\documentclass[5p,authoryear,10pt]{elsarticle}

\usepackage{amssymb}
\usepackage{amsmath}
\usepackage{natbib}
\usepackage{subfigure}
\usepackage{graphicx}
\usepackage{epstopdf}
\usepackage{marvosym}
\usepackage{epsfig}
\usepackage{multirow}
\usepackage{tikz}
\usepackage{hyphenat} 
\usepackage{array} 
\usetikzlibrary{arrows,calc,intersections,positioning,arrows,decorations.pathmorphing,decorations.markings,shapes}%

\usepackage{tikz-3dplot}
\newcolumntype{$}{>{\global\let\currentrowstyle\relax}}
\newcolumntype{^}{>{\currentrowstyle}}

\journal{Planetary and Space Science}

\begin{document}

\begin{frontmatter}

\title{On the contribution of PRIDE-JUICE to Jovian system ephemerides}

%% TO DO %%%%%%%%%%%%%%%%%%%%%%%%%%%%%%%%%%%%%%%%%%%%%%%%%%%%%%%%%%%%%%%
%%
%%   Giuseppe Table 5 comment: add + sign for positive change (i.e worsening results)
%%   Check Europa improvement result

\author[JIVE,TUDELFT]{D. Dirkx\corref{cor}}
\cortext[cor]{Corresponding author; Tel.: +31(0)15 2788866; Fax: +31(0)15 2781822 }
\ead{D.Dirkx@tudelft.nl}
\author[JIVE,TUDELFT]{L.I. Gurvits}
\ead{LGurvits@jive.eu}
\author[IMCCE]{V. Lainey}
\ead{Valery.Lainey@imcce.fr}
\author[PISA]{G. Lari}
\ead{lari@student.dm.unipi.it}
\author[PISA]{A. Milani}
\ead{milani@dm.unipi.it}
\author[JIVE,ASTRON]{G. Cim\`{o}}
\ead{cimo@jive.eu}
\author[JIVE,TUDELFT,SHANGHAI]{T.M. Bocanegra-Bahamon}
\ead{T.M.BocanegraBahamon@tudelft.nl}
\author[TUDELFT]{P.N.A.M. Visser}
\ead{P.N.A.M.Visser@tudelft.nl}

\address[JIVE]{Joint Institute for VLBI ERIC, PO Box 2, 7990 AA Dwingeloo, The Netherlands}
\address[TUDELFT]{Delft University of Technology, Kluyverweg 1, 2629HS Delft, The Netherlands}
\address[IMCCE]{IMCCE, Observatoire de Paris, PSL Research University, CNRS-UMR8028 du CNRS, UPMC, Lille-1, 77 Av. Denfert-Rochereau, 75014, Paris, France}
\address[PISA]{Department of Mathematics, University of Pisa, via Buonarroti 2, 56127, Pisa, Italy}
\address[ASTRON]{ASTRON, the Netherlands Institute for Radio Astronomy, Postbus 2, 7990 AA, Dwingeloo, The Netherlands}
\address[SHANGHAI]{Shanghai Astronomical Observatory, 80 Nandan Road, 200030 Shanghai, PR China}

%%%%%%%%%%%%%%%%%%%%%%%%%%%%%%%% TO DO%%%%%%%%%%%%%%%%%%%%%%
% Add reference for Thebe mass in Section 3.1
% Add angle \psi to Fig. 1
\begin{abstract}
{The Jupiter Icy Moons Explorer (JUICE) mission will perform detailed measurements of the properties of the Galilean moons, with a nominal {in-system} science-mission duration of about 3.5 years. Using both the radio tracking data, and (Earth- and JUICE-based) optical astrometry, the dynamics of the Galilean moons will be measured to unprecedented accuracy. This will provide crucial input to the determination of the ephemerides and physical properties of the system, most notably the dissipation in Io and Jupiter.}

{The data from Planetary Radio Interferometry and Doppler Experiment (PRIDE) will provide the lateral position of the spacecraft in the International Celestial Reference Frame (ICRF). In this article, we analyze the relative quantitative influence of the JUICE-PRIDE observables to the determination of the ephemerides of the Jovian system and the associated physical parameters. }
{We perform a covariance analysis for a broad range of mission and system characteristics. We analyze the influence of VLBI data quality, observation planning, as well as the influence of JUICE orbit determination quality. This provides key input for the further development of the PRIDE observational planning and ground segment development.}

 {Our analysis indicates that the VLBI data are especially important for constraining the dynamics of Ganymede and Callisto perpendicular to their orbital planes. Also, the use of the VLBI data makes the uncertainty in the ephemerides less dependent on the error in the orbit determination of the JUICE spacecraft itself. Furthermore, we find that optical astrometry data of especially Io using the JANUS instrument will be crucial for stabilizing the solution of the normal equations. Knowledge of the dissipation in the Jupiter system cannot be improved using satellite dynamics obtained from JUICE data alone, the uncertainty in Io's dissipation obtained from our simulations is similar to the present level of uncertainty.}
\end{abstract}


\end{frontmatter}

\section{Introduction}
\label{sec:introduction}

The Jupiter Icy Moons Explorer (JUICE) mission will study the Jovian system in 2030-2033, with a focus on investigating the icy moons Europa, Ganymede and Callisto \citep{GrassetEtAl2013,TitovEtAl2014}. To attain its science goals, the mission carries 11 scientific experiments. In this article, we focus on the use of the Planetary Radio Interferometry and Doppler Experiment (PRIDE) \citep{GurvitsEtAl2013}. The PRIDE experiment is unique, in the sense that it requires no dedicated on-board hardware beyond what will already be available aboard the JUICE spacecraft for communications, tracking, and the Gravity and Geophysics of Jupiter and the Galilean Moons (3GM) experiment \citep{Iess2013}.

\begin{figure*}[tbp!]
\centering
\includegraphics[width=0.97\textwidth]{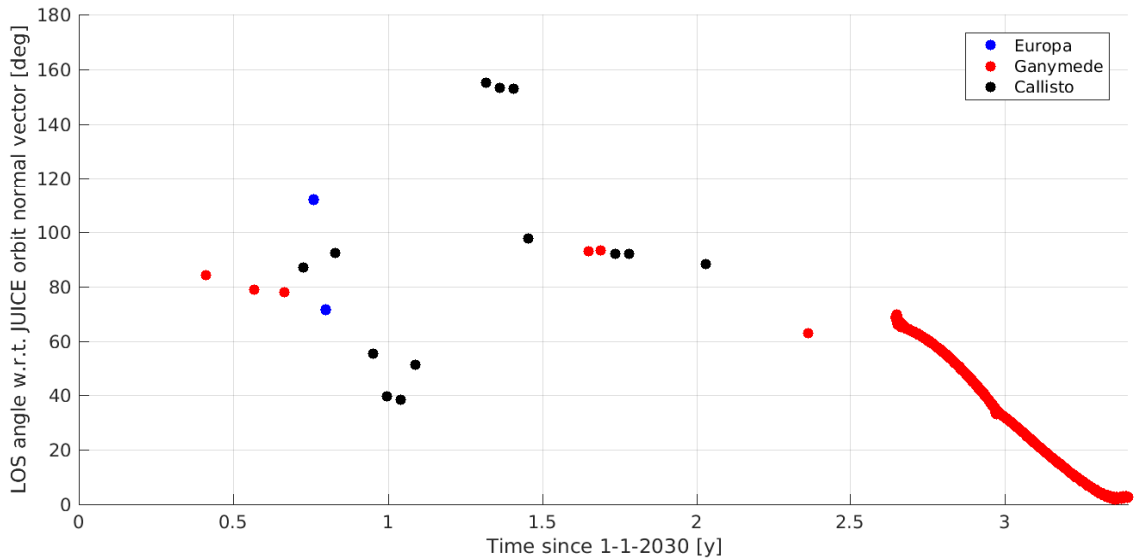}
\caption{Angle between the line-of-sight (LOS) vector from Earth, and the orbital plane of the spacecraft w.r.t. the satellites during the flybys/orbit phase. Angles are plotted while the spacecraft is inside the sphere of influence of the satellites. An angle of 90 degrees represents edge-on viewing; 0 and 180 degrees represnet face-on viewing.}
\label{fig:flybyGeometry}
\end{figure*}

The mission profile consists of a number of flybys of Ganymede and Callisto over a period of 2.5 years, as well as two flybys of Europa spaced two weeks apart. Subsequently, the spacecraft will enter orbit around Ganymede, initially with a semi-major axis of about 8,000 km during the GCO/ GEO5000 phase (eccentricity $e$ between 0.0 and 0.6) lasting about 5 months. Thereafter, the spacecraft will be inserted into its final spherical orbit at 500 km altitude (denoted GCO500), which is to last for a nominal duration of 4 months. The JUICE mission design that we have used here is provided in the {Consolidated Report on the Mission Analysis (CREMA), version 2.0. The timing of the flybys, as well as the viewing geometry, is shown in Fig. \ref{fig:flybyGeometry}}.

% Initially, the orbit will have a semi-major axis of around 8,000 km. During this initial phase, the spacecraft will fly in both a circular and an elliptical orbit (eccentricity $e$ up to 0.6), as a natural result of Jupiter's perturbations of its orbit (this phase is denoted GCO/GEO5000). After a period of about 4 months, the spacecraft will be inserted into its final spherical orbit at 500 km altitude (denoted GCO500). The nominal mission calls for a 4 month 500 km orbit phase, with any possible mission extension likely either continuing in this orbit, or proceeding to a lower orbit. %It is noteworthy that the original mission called for a 200 km altitude phase following the 500 km phase. However, this low orbit was scrapped for budgetary reasons.

%\begin{figure*}[tbp!]
%\centering
%\includegraphics[width=0.77\textwidth]{Figures/JuiceFlybyGeometry.eps}
%\caption{Observational geometry of the moons' orbits during the flybys.}
%\label{fig:flybyGeometry}
%\end{figure*}

The PRIDE experiment uses the radio signals transmitted by the spacecraft's High and Medium Gain Antennas (HGA and MGA) to determine the lateral position of the spacecraft, \emph{i.e.,} the position in the two mutually orthogonal directions perpendicular to the line-of-sight vector \citep{DuevEtAl2012} in the International Celestial Reference Frame (ICRF). These measurements are obtained by means of Very Long Baseline Interferometry (VLBI), using a large ($\gtrsim$10) number of Earth-based radio telescopes to simultaneously observe the radio signal emitted by the spacecraft. In addition to the lateral position observable, the PRIDE experiment produces \emph{ad hoc} Doppler data of the spacecraft's radial velocity, as observed by each ground station \citep{DuevEtAl2016}. 

The PRIDE data can be used to improve the frame ties between the dynamical reference frame and the ICRF \citep{ParkEtAl2015b}, to probe the Jovian atmosphere by means of radio occultation observations \citep{BocanegraBahamonEtAl2015}, to complement the Doppler data from 3GM for JUICE orbit determination, and to improve the ephemerides of the Jovian system \citep{GurvitsEtAl2013}. {Also, for periods of poor observational geometry (see Fig. \ref{fig:flybyGeometry}), the VLBI data itself may be used to better constrain the spacecraft's orbit.} Our focus in this article is an analysis of the improvement in the ephemerides, in particular for the Galilean moons. Especially the orbital phase of JUICE at Ganymede will allow the ephemerides to be improved dramatically. JUICE orbit determination uncertainty in this phase will be $<10$ m \citep{ParisiEtAl2012}, allowing the signature of the satellite's dynamics to be much more accurately extracted from the tracking data than is the case for flybys. Moreover, the fact that the motion of Ganymede will be measured over dozens of full orbital periods (indirectly, through JUICE), as opposed to snapshots provided by flybys, will allow the ephemerides to be much more accurately determined than has been the case for planetary satellites thus far.

VLBI measurements have been used on a large number of past and current planetary missions such as Vega \citep{PrestonEtAl1986}, Ulysses \citep{FolknerEtAl1996} Cassini-Huygens \citep{PogrebenkoEtAl2004,LebretonEtAl2005,JonesEtAl2015}, Chang'E-1 \citep{JianguoEtAl2010}, IKAROS \citep{TakeuchiEtAl2011}, Venus Express \citep{DuevEtAl2012}, MRO \citep{ParkEtAl2015b}, Mars Express \citep{DuevEtAl2016} and Juno \citep{JonesEtAl2017}. Present state-of-the-art measurements provide lateral position (right ascension $\alpha$ and declination $\delta$) measurements with an uncertainty in ICRF of approximately 1.0 nrad ($\approx$ 200 $\mu$as) or slightly better. 

The range measurements from 3GM, supplemented by the lateral position data from PRIDE, will be invaluable for the construction of improved ephemerides of Jupi\-ter and its moons. Doppler data, produced by 3GM and supplemented by PRIDE, will be the primary input for the determination of the spacecraft's trajectory, as is typical in space mission tracking data analysis \citep[\textit{e.g., }][]{PitjevaEtAl2001,FiengaEtAl2014,JonesEtAl2015}. The Doppler data will have negligible impact on the determination of the planetary ephemeris, but does contain crucial information on the satellite dynamics, in addition to the spacecraft orbit.

{Improved satellite ephemerides will benefit a range of scientific fields}. Various properties of the Galilean moons and Jupiter can potentially be constrained by reconstructing the moons' dynamics \citep{LaineyEtAl2009, DirkxEtAl2016}. {In particular the determination of tidal dissipation in the system can be uniquely constrained by accurately reconstructing the bodies' dynamics over long time intervals ($>$100 years)} The dynamical behaviour of the Galilean moons is crucial in understanding the long-term behaviour of the Jovian system \citep{Peale1999,Greenberg2010}, as well as the internal processes of the moons \citep{HussmannEtAl2010}. {Furthermore, the present configuration and properties of the Galilean moons can be used to shed insight onto the formation and evolution of the solar system \citep{DeiennoEtAl2014,HellerEtAl2015}. Additionally, the investigation of the Galilean moons (most notably Europa) as a possible habitat \citep{MarionEtAl2003} would greatly benefit from an improved determination of both the internal heat production and long-term thermal-orbital evolution of the moons. From a more practical point of view, improving satellite ephemerides allows spacecraft to target flybys, or orbit insertions, more accurately \citep{MurrowJacobson1988}. This results in a more robust mission planning, and a reduced $\Delta V$ budget allocated to orbit corrections, reducing the mass and cost of a mission.}

A sensitivity analysis of the dynamics of the Galilean moons to various physical characteristics (\emph{i.e.,} tidal, rotational and gravitational) was performed by \cite{DirkxEtAl2016}, with a focus on the JUICE mission. There, it is concluded that the influence of most physical parameters will be absorbed into the estimation of the moons' initial states. Important exceptions are the $k_{2}/Q$ of Io and Jupiter, which have been previously estimated by \cite{LaineyEtAl2009} using 117 years of astrometry data. \cite{DirkxEtAl2016} also conclude that Europa's $k_{2}/Q$ and Io's $k_{2}$, which are both at present unconstrained from observations of the dynamics, may be observable.

Our goal in this article is to analyze the relative contribution of the lateral position (VLBI) data obtained from the PRIDE experiment.  We determine the formal uncertainty of the position of the Galilean moons, and the associated physical parameters, from simulated radio tracking and optical astrometry data that will be collected by the JUICE mission, providing insight into the science return of the mission, and the synergy between the various experiments. The results presented here will be crucial in the further development of both the PRIDE observational planning, as well as ground segment infrastructure and data analysis upgrades.  We vary a broad range of system and mission characteristics to analyze the influence of the experiment under many different conditions, providing a clear and quantitative picture of the strengths and weaknesses of PRIDE's VLBI data. 
%Our goal is to study the \emph{relative} contribution of the PRIDE observables for a broad range of characteristics. 
As we aim to analyze the influence of a large number of combinations of characteristics, we require a model that is computationally efficient in running a single simulation. To achieve the required computational efficiency and make our analysis tractable, we decouple the orbit determination of the spacecraft from the ephemeris generation. 

We outline the methods and models for the simulation of the tracking data in Section \ref{sec:dataModelling}, where we discuss the decoupling between ephemeris generation and orbit determination and the impact this has on our results. Models used to simulate the ephemeris uncertainty are presented in Section \ref{sec:ephemerisMethodology}. The results of analysis of the Galilean satellite ephemeris uncertainty are discussed in Section \ref{sec:results}. The results for the uncertainty of the Jovian ephemeris and physical parameters of the Jovian system are discussed in Section \ref{sec:jovianSystemResults}. Finally, we present our conclusions in Section \ref{sec:conclusions}.

%\begin{figure*}[tbb!]
%\centering
%\includegraphics[width=0.97\textwidth]{Figures/JuiceGanymedeOrbit.eps}
%\caption{Orbital elements ofJUICE during the Ganymede orbit phase.}
%\label{fig:ganymedeOrbitElements}
%\end{figure*}

%\section{Methodology}
%\label{sec:methodology}
%In this section, we describe the models we use to simulate the uncertainty of the ephemerides of the Jovian system from JUICE tracking data. %In Section \ref{sec:juicePlanning}, we provide an overview of the aspects of the JUICE mission relevant for this study, followed by a detailed discussion of the tracking data, the consideration of the JUICE spacecraft and the weights we use in the simulations in Sections \ref{sec:trackingData}, \ref{sec:juiceDynamics} and \ref{sec:dataWeight}, respectively. In Section \ref{sec:estimatedParameters}, we discuss the parameters that we use in our simulations. We make use of covariance analysis, the details for which are discussed in Section \ref{sec:covarianceAnalysis}. The observation scheduling that we use in our simulations is discussed in Sections \ref{sec:obsPlanning}. Finally, we discuss the inherent rank deficiency in our estimation problem, as well as mitigation strategies, in Section \ref{sec:aprioriInformation}.

\section{Data modelling and simulation}
\label{sec:dataModelling}
In this section, we discuss the JUICE tracking data that we use in our simulations. We start by discussing the types of tracking data in Section \ref{sec:trackingData}. We discuss considerations related to orbit determination of the JUICE spacecraft in Section \ref{sec:juiceDynamics} and present the weights we use for the data in Section \ref{sec:dataWeight}. 

\begin{figure*}[htb!]
\centering
\tdplotsetmaincoords{60}{110}

%define polar coordinates for some vector
%TODO: look into using 3d spherical coordinate system

\pgfmathsetmacro{\rvecJup}{7}
\pgfmathsetmacro{\thetavecJup}{45}
\pgfmathsetmacro{\phivecJup}{100}

\pgfmathsetmacro{\rvecmoon}{13}
\pgfmathsetmacro{\thetavecmoon}{55}
\pgfmathsetmacro{\phivecmoon}{70}

\pgfmathsetmacro{\xmoon}{\rvecmoon*sin(\thetavecmoon)*cos(\phivecmoon)}
\pgfmathsetmacro{\ymoon}{\rvecmoon*sin(\thetavecmoon)*sin(\phivecmoon)}
\pgfmathsetmacro{\zmoon}{\rvecmoon*cos(\thetavecmoon)}

\pgfmathsetmacro{\rvecsc}{8}
\pgfmathsetmacro{\thetavecsc}{65}
\pgfmathsetmacro{\phivecsc}{80}

\pgfmathsetmacro{\xsc}{\rvecsc*sin(\thetavecsc)*cos(\phivecsc)}
\pgfmathsetmacro{\ysc}{\rvecsc*sin(\thetavecsc)*sin(\phivecsc)}
\pgfmathsetmacro{\zsc}{\rvecsc*cos(\thetavecsc)}

\pgfmathsetmacro{\xmoonLocal}{\xmoon-\xsc}
\pgfmathsetmacro{\ymoonLocal}{\ymoon-\ysc}
\pgfmathsetmacro{\zmoonLocal}{\zmoon-\zsc}

\pgfmathsetmacro{\rmoonLocal}{sqrt(\xmoonLocal^2+\ymoonLocal^2+\zmoonLocal^2)}
\pgfmathsetmacro{\thetamoonLocal}{acos(\zmoonLocal/\rmoonLocal)}
\pgfmathsetmacro{\phimoonLocal}{atan(\ymoonLocal/\xmoonLocal)}
%start tikz picture, and use the tdplot_main_coords style to implement the display 
%coordinate transformation provided by 3dplot
\begin{tikzpicture}[scale=1.3,tdplot_main_coords]

%set up some coordinates 
%-----------------------
\coordinate (O) at (0,0,0);

%draw the main coordinate system axes
\draw[thick,->] (0,0,0) -- (5,0,0) node[anchor=north east]{$x$};
\draw[thick,->] (0,0,0) -- (0,9,0) node[anchor=north west]{$y$};
\draw[thick,->] (0,0,0) -- (0,0,5) node[anchor=south]{$z$};

%determine a coordinate (P) using (r,\theta,\phi) coordinates.  This command
%also determines (Pxy), (Pxz), and (Pyz): the xy-, xz-, and yz-projections
%of the point (P).
%syntax: \tdplotsetcoord{Coordinate name without parentheses}{r}{\theta}{\phi}
\tdplotsetcoord{Pmoon}{\rvecmoon}{\thetavecmoon}{\phivecmoon}
\tdplotsetcoord{Psc}{\rvecsc}{\thetavecsc}{\phivecsc}
\tdplotsetcoord{PJup}{\rvecJup}{\thetavecJup}{\phivecJup}

\fill (O) circle [fill=black,radius=0.15cm] node[left=0.2cm of O] {\small Earth  };
\fill (Pmoon) circle [fill=black,radius=0.15cm] node[below right] {\small Satellite $i$  };
\fill (Psc) circle [fill=black,radius=0.15cm] node[above left] {\small JUICE };
\fill (PJup) circle [fill=black,radius=0.15cm] node[left=0.2 cm of PJup] {\small Jupiter  };

%draw a vector from origin to point (P) 
\draw[thin,-latex,color=red] (O) -- node[below]{$\mathbf{r}_{i}$} (Pmoon);

%draw projection on xy plane, and a connecting line
\draw[dashed, color=blue] (O) -- (Pmoonxy);
\draw[dashed, color=blue] (Pmoon) -- (Pmoonxy);
\draw[dotted,color=blue] (\xmoon,0,0)-- (\xmoon,\ymoon,0);
\draw[dotted,color=blue] (0,\ymoon,0)-- (\xmoon,\ymoon,0);

%draw a vector from origin to point (P) 
\draw[-latex,color=red] (O) -- node[below]{$\mathbf{r}_{sc}$} (Psc) ;
\draw[-latex,color=red] (O) -- node[right]{$\mathbf{r}_{J}$} (PJup) ;
\draw[-latex,color=red] (PJup) -- node[below]{$\mathbf{r}_{i}^{J}$} (Pmoon) ;

%draw projection on xy plane, and a connecting line
\draw[dashed, color=blue] (O) -- (Pscxy);
\draw[dashed, color=blue] (Psc) -- (Pscxy);
\draw[dotted,color=blue] (\xsc,0,0)-- (\xsc,\ysc,0);
\draw[dotted,color=blue] (0,\ysc,0)-- (\xsc,\ysc,0);

%draw the angle \phi, and label it
%syntax: \tdplotdrawarc[coordinate frame, draw options]{center point}{r}{angle}{label options}{label}
\tdplotdrawarc[-latex]{(O)}{1.0}{0}{\phivecmoon}{anchor=north}{$\alpha_{i}$}

%set the rotated coordinate system so the x'-y' plane lies within the
%"theta plane" of the main coordinate system
%syntax: \tdplotsetthetaplanecoords{\phi}
\tdplotsetthetaplanecoords{\phivecmoon}

%draw theta arc and label, using rotated coordinate system
\tdplotdrawarc[latex-,tdplot_rotated_coords]{(0,0,0)}{1.0}{\thetavecmoon}{90}{anchor=south west}{$\delta_{i}$}

%draw the angle \phi, and label it
%syntax: \tdplotdrawarc[coordinate frame, draw options]{center point}{r}{angle}{label options}{label}
\tdplotdrawarc[-latex]{(O)}{2}{0}{\phivecsc}{anchor=north}{$\alpha_{sc}$}
\tdplotsetthetaplanecoords{\phivecsc}
\tdplotdrawarc[tdplot_rotated_coords,latex-]{(0,0,0)}{2}{\thetavecsc}{90}{anchor=south west}{$\delta_{sc}$}

\draw[latex-,color=red] (Psc) -- node[right]{$\mathbf{r}_{sc}^{i}$} (Pmoon);

\tdplotsetrotatedcoords{0}{0}{0}
\tdplotsetrotatedcoordsorigin{(Psc)}

\draw[dashed,color=blue,tdplot_rotated_coords,-] (0,0,0)-- (\xmoonLocal,\ymoonLocal,0);
\draw[dotted,color=blue,tdplot_rotated_coords,-] (\xmoonLocal,0,0)-- (\xmoonLocal,\ymoonLocal,0);
\draw[dotted,color=blue,tdplot_rotated_coords,-] (0,\ymoonLocal,0)-- (\xmoonLocal,\ymoonLocal,0);

\tdplotdrawarc[tdplot_rotated_coords,-latex]{(0,0,0)}{1.5}{0}{\phimoonLocal}{anchor=north}{$\alpha_{i}^{sc}$}
\tdplotsetthetaplanecoords{\phimoonLocal}
\tdplotdrawarc[tdplot_rotated_coords,latex-]{(0,0,0)}{1.5}{\thetamoonLocal}{90}{anchor=west}{$\delta_{i}^{sc}$}

\tdplotsetrotatedcoords{0}{0}{0}
\tdplotsetrotatedcoordsorigin{(Psc)}

\draw[thick,color=black,tdplot_rotated_coords,->] (0,0,0)
-- (3,0,0) node[anchor=south]{$x^{sc}$};
\draw[thick,color=black,tdplot_rotated_coords,->] (0,0,0)
-- (0,3,0) node[anchor=west]{$y^{sc}$};
\draw[thick,color=black,tdplot_rotated_coords,->] (0,0,0)
-- (0,0,3.3) node[anchor=south]{$z^{sc}$};

\tdplotsetrotatedcoords{0}{0}{0}
\tdplotsetrotatedcoordsorigin{(O)}

\end{tikzpicture}
\caption{Schematic representation of relevant positions and observables for the JUICE mission. {Vectors $\mathbf{r}_{k}$ denote the position of element $k$ w.r.t. Earth, while $\mathbf{r}_{k}^{i}$ denotes their position w.r.t. body $i$. Here, indices $J$, $i$ and $sc$ denote Jupiter, satellite $i$ (Io=1, Europa=2, Ganymede=3, Callisto=4) and the spacecraft, respectively. Similarly, $\alpha_{k}$ and $\delta_{k}$ denote right ascension and declination of element $k$, where no superscript denotes a measurement from Earth, and $sc$ superscript the measurement from the spacecraft.  Note that the all reference frame have the same orientation as the global frame. }}
\label{fig:observableGeometry}
\end{figure*}
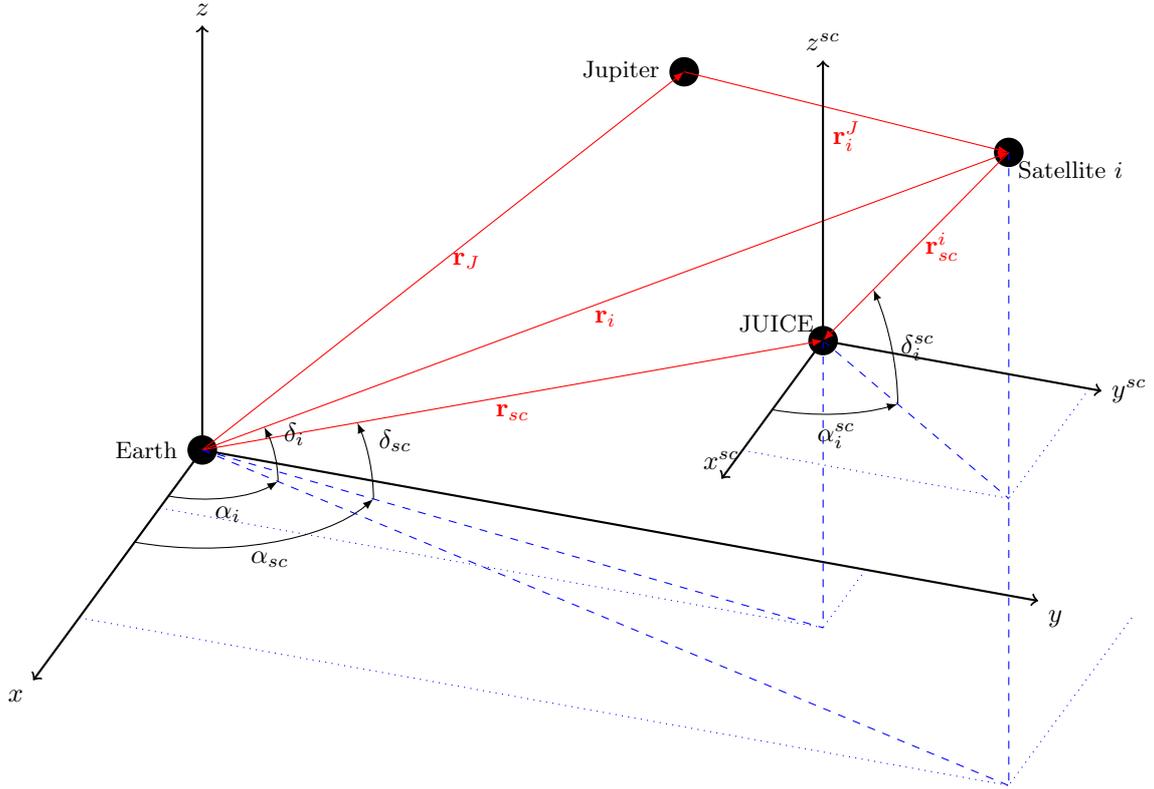
\subsection{Tracking Data Type Selection}
\label{sec:trackingData}

%In this section, we present the types of the observables that we use in our simulations. 
%Our goal in this article is to identify the contribution that the PRIDE data will have on the science goals of the JUICE mission, in particular the determination of satellite ephemerides and associated physical parameters. Although the \emph{ad hoc} Doppler data that PRIDE will produce may provide a supplementary role to the 3GM data, we focus here on the VLBI observable, as this constitutes the main product of the experiment.
The information content of the moon dynamics is encoded in the 3GM and PRIDE data (Doppler, range and VLBI) by its signature on the dynamics of the spacecraft when it performs a flyby, or is in orbit of, one of the moons \citep[\emph{e.g.}][]{Moyer2005,MilaniGronchi2010}. 

%The closed-loop Doppler observable (which we denote as $\dot{r}_{sc}(t)$), which is typically realized by dedicated tracking stations, such as those of Estrack and  the Deep Space Network (DSN), are generated from an accumulated cycle count of the electromagnetic signal at the ground station. This observable can be modelled as two range measurements ${r}_{sc}(t)$, differenced over an integration time $\Delta t$, so \citep{Moyer2005}:
%\begin{align}
%\dot{r}_{sc}(t)=\frac{{r}_{sc}(t)-{r}_{sc}(t-\Delta t)}{\Delta t}\label{eq:dopplerDataType}
%\end{align}
%Therefore, the type of information content of the range and closed-loop Doppler observables is comparable, with the key difference being that Doppler data can much more accurately measure changes in range over short time periods (on the order of ten hours) than the range data. Range data is expected to have an uncertainty of 0.2 m, compared to 0.01 mm/s for Doppler data over 60 s integration time \citep{Iess2013}. We note that Doppler measurements by PRIDE employ a different approach of direct analysis of the radio signal spectrum as discussed by \cite{DuevEtAl2012,DuevEtAl2016}, generating so-called open-loop Doppler data. A comparison and possible combination of the Doppler measurement sets obtained by 3GM and PRIDE is outside of the scope of this paper and will be discussed elsewhere.  

The dynamics of the spacecraft w.r.t. the moons (position and velocity denoted by $\mathbf{r}_{sc}^{i}(t)$ and $\dot{\mathbf{r}}_{sc}^{i}(t)$, respectively, for moon $i$) is reconstructed almost entirely from the Doppler data. %, which is expected to be at the 10 $\mu$m/s level for integration times on the order of 60 s \citep{Iess2013}. 
Here, we are only indirectly interested in the orbit determination of the spacecraft itself, through its coupling with the ephemeris generation. 
%Moreover, as we are interested in the \emph{relative} contribution of PRIDE-JUICE, our requirements for the absolute accuracy of the uncertainties of the ephemerides that we simulate are relatively loose. 
Including the estimation of the spacecraft dynamics directly in our simulations would greatly increase the runtime of a single simulation, as it would require a (constrained) multi-arc solution \citep[\emph{e.g.}][]{AlessiEtAl2012} of the spacecraft over the full mission duration. For our simulations, we employ covariance analyses of the estimation quality. Covariance analysis for spacecraft orbit determination is typically much more optimistic (in terms of true-to-formal error ratio) than for planetary/satellite ephemerides (Section \ref{sec:covarianceAnalysis}). Coupling the analysis of the two would complicate the interpretation of the formal errors that we obtain for the satellites. 

For these reasons, we do not {directly} estimate the state of the spacecraft. However, the orbit determination of the spacecraft is indirectly included, as discussed in detail in Section \ref{sec:juiceDynamics}. By decoupling the orbit determination from the ephemeris generation and introducing spacecraft orbit errors as mission settings (see Section \ref{sec:dataWeight}), we parametrically include the influence of orbit determination quality. This allows us to perform our broad analysis, while still retaining sufficient model fidelity to analyze the relative influence of JUICE-PRIDE data.

The Doppler data is by far the main contributor to the JUICE orbit determination. Nevertheless, it is also crucial for the ephemeris generation, especially during the flybys, and we therefore retain it in our analyses. The data types we use for our simulations are:
\begin{itemize}
\item Range $|\mathbf{r}_{sc}|(=r_{sc})$ from Earth-based tracking stations to the spacecraft during flybys/orbit of the moons, obtained by 3GM
\item Doppler $|\dot{\mathbf{r}}_{sc}|(=\dot{r}_{sc})$ from Earth-based tracking stations to the spacecraft during flybys/orbit of the moons, obtained by primarily by 3GM, and supplemented by PRIDE\footnote{Note that Doppler data from 3GM is obtained in a closed-loop mode (as opposed to open-loop mode of PRIDE), in which the observed range-rate $\dot{r}_{sc}$ is averaged over some time $\Delta t$, which is typically in the range of 1-1000 s. For the purposes of our simulation study, the information content encoded in the instantaneous and averaged Doppler data is qualitatively equivalent, though}.
\item Lateral positions (VLBI data) observed from Earth, referenced to the geocenter, ($\delta_{sc}, \alpha_{sc}$) of the spacecraft during flybys/orbit of the moons, obtained by PRIDE
\item Optical astrometry of the moons $i$ from the {JANUS instrument and/or the navigation camera (NavCam)} ($\delta_{i}^{sc}$, $\alpha_{i}^{sc}$)
\item Optical astrometry of the moons $i$ from Earth-based observatories ($\delta_{i}, \alpha_{i}$)
\end{itemize}
These quantities are shown schematically in Fig. \ref{fig:observableGeometry}. The first three tracking types are observations of the JUICE spacecraft, whereas the latter two are direct, optical astrometric, observations of the moons. Both spacecraft- and Earth-based optical astrometry have been used in the generation of the ephemerides of various planetary satellites \citep[\emph{e.g.}][]{LaineyEtAl2007,RosenblattEtAl2013,LaineyEtAl2015}.

We note that the generation of satellite epheme\-rides from JUICE data will be fundamentally different than has been the case for other missions (\emph{e.g.} Galileo, Cassini), due to the orbit phase at Ganymede. This will allow the motion of Ganymede to be sampled over many orbits, as opposed to the snapshot positions that are obtained from flybys. The full set of tracking data during the orbit phase will contain a superposition of numerous signatures, primary among them the orbit of Ganymede around Jupiter (period of about 172 hours) and JUICE around Ganymede (about 3 hours). The 172 hour signature in the data, from which Ganymede's ephemeris will be largely obtained, makes the range data much more influential than is the case for flybys (where Doppler data is by far the primary source of information for satellite ephemerides). In fact, analyses by \cite{Dirkx2015,CicaloEtAl2016} show that 10 $\mu$m/s Doppler data (at 60 s integration time) will be less sensitive than 20 cm range data to a signature of 172 hours. \cite{CicaloEtAl2016} analyze the performance of the BepiColombo radio tracking system, which is similar to that which will be carried by JUICE, and conclude that the signature of effects with a period $>10^5$ s ($\approx$28 hours) are better observed in the range data, compared to the Doppler data.

%\begin{figure*}[tbb!]
%\centering
%\includegraphics[width=0.97\textwidth]{Figures/JuiceGanymedeOrbit.eps}
%\caption{Orbital elements ofJUICE during the Ganymede orbit phase.}
%\label{fig:ganymedeOrbitElements}
%\end{figure*}

\subsection{Consideration of JUICE orbit uncertainty}
\label{sec:juiceDynamics}
As discussed in the previous section, we omit the orbit determination of the spacecraft from our simulations. In this section, we will discuss the consequences of this assumption, and indicate how we indirectly include it in a parametric manner for our simulations.

The range and VLBI observables encode the three components of the vector:
\begin{align}
\mathbf{r}_{sc}=\mathbf{r}_{sc}^{I}-\mathbf{r}_{E}^{I}\label{eq:measurementComposition}
\end{align}
while the Doppler data encodes the time-derivative of its line-of-sight component. Here, the superscript denotes the origin of the frame in which the vector is expressed ($I$ denotes an inertial barycentric frame), and the subscript denotes the body of which the position is expressed ($E$ denotes Earth). However, in our analysis, we are interested in the state of the moon $i$ at which the flyby occurs, w.r.t. Jupiter, denoted $\mathbf{r}_{i}^{J}$.  %Consequently, the physical signal we wish to retrieve is encoded in the term ($\mathbf{r}_{i}^{I}=\mathbf{r}_{J}^{I}$+$\mathbf{r}_{i}^{J}$). 
To relate this term to (the components of) our observed $\mathbf{r}_{sc}$, we decompose the spacecraft position as:
\begin{align}
\mathbf{r}_{sc}^{I}=\mathbf{r}_{sc}^{i}+\mathbf{r}_{i}^{J}+\mathbf{r}_{J}^{I}\label{eq:measurementComposition2}
\end{align}
which includes the JUICE spacecraft position w.r.t. the moon, denoted $\mathbf{r}_{sc}^{i}$, the barycentric position of Jupiter $\mathbf{r}_{J}^{I}$, as well as the dynamics $\mathbf{r}_{i}^{J}$ we wish to observe. Both $\mathbf{r}_{J}^{I}$ and $\mathbf{r}_{i}^{J}$, for $i=1..4$, are included in our estimation from simulated tracking data, while $\mathbf{r}_{sc}^{i}$ is not.

%From the above, the total uncertainty of the measurement can be decomposed as:
%\begin{align}
%\sigma\left({\mathbf{r}_{sc}}\right)=\sigma\left({\mathbf{r}_{sc}^{i}}+{\mathbf{r}_{i}^{J}}+\left({\mathbf{r}_{J}^{I}}-{\mathbf{r}_{E}^{I}}\right)\right)
%\end{align}
%Since we do not estimate the orbit of the JUICE spacecraft in our simulations, its uncertainty will be introduced in the effective uncertainty of the range between the ground station and the target moon. 

Since we  decouple our analysis of the ephemeris uncertainty from the orbit determination of the spacecraft, we are left with the challenge of properly including the uncertainties of the determination of $\mathbf{r}_{sc}^{i}$ into our analysis. There is a strong coupling between the orbit determination of the spacecraft during a flyby, and the estimated Jupiter-centered position of the moon during that flyby. Orbit determination of the spacecraft, in which the uncertainty in the moons' dynamics is not considered, will only yield a conditional uncertainty of the spacecraft's dynamics. Using such a conditional orbit determination during the ephemeris generation will lead to optimistic results (in terms of formal errors), as the conditional uncertainty of the satellites' states is much smaller than its marginal uncertainty \citep{MilaniGronchi2010}. We note that this effect will be especially pronounced during the flybys. 

We neglect the Earth's ephemeris uncertainty (compared to that of Jupiter), and we can write the conditional uncertainty of the signature of the moon and planetary dynamics in the measurements as:
\begin{align}
\sigma\left({\mathbf{r}_{i}^{J}}+{\mathbf{r}_{J}^{I}}\right)|_{\mathbf{r}_{sc}^{i}=\text{fixed}}=\sigma\left({\mathbf{r}_{sc}}\right)+\sigma\left({\mathbf{r}_{sc}^{i}}\right)|_{\mathbf{r}_{i}^{J}=\text{fixed}}\label{eq:conditionalMeasurementUncertainty}
\end{align}
where $\mathbf{r}_{sc}$ is defined by Eq. (\ref{eq:measurementComposition}). We see that the conditional uncertainty in the measurements $\mathbf{r}_{i}^{I}(=\mathbf{r}_{J}^{I}$+$\mathbf{r}_{i}^{J}$) is the combination of the measurement uncertainty and JUICE's position uncertainty w.r.t. the moon under consideration. 

Ideally, we would estimate the dynamics of the spacecraft in a single arc during the mission, concurrently and consistently estimating it with the moons. However, this is both practically and fundamentally not possible. Practically, errors in the dynamical modelling of the spacecraft will prevent the accurate reconstruction of the dynamics over such long time periods (even when estimating corrections to force models during the inversion). Fundamentally, the complete dynamics of the spacecraft over the course of its $\sim\,20$ flybys will be chaotic, so that the observations span beyond the computability horizon \citep{SpotoMilani2016}, making the use of a single arc impossible. This issue could be tackled by the use of the constrained multi-arc methodology \citep{AlessiEtAl2012,CicaloEtAl2016}. Using this method, information is passed between neighbouring arcs in the form of constraints, recognizing the fact that the estimated arcs belong to the same object, while preventing a single arc from passing beyond the computability horizon. Such an approach would be crucial in the robust and detailed analysis of the spacecraft orbit determination.
%Although the combination of the constrained multi-arc estimation of JUICE, and ephemeris generation of the moons would provide a more elegant and rigorous analysis of the potential science return from the full suite of JUICE tracking data, 
Such an approach is not feasible here, however, as we must analyze many $(\sim 10^{5})$ different cases of mission and system parameters, requiring a fast model for a single evaluation. Moreover, our primary goal is not to provide absolute uncertainties, but instead to provide insight into the {relative} contribution of PRIDE-JUICE. 

When using Eq. (\ref{eq:conditionalMeasurementUncertainty}) to obtain uncertainties for measurements of satellite state, the main model simplification that needs to be overcome is the implicit assumption that the conditional uncertainties of $\mathbf{r}_{i}^{J}$ and $\mathbf{r}_{sc}^{i}$ are the same as their marginal uncertainties. For the orbital phase, this assumption is reasonable, as the signature of Ganymede's orbit about Jupiter will be distinct from JUICE's orbit about Ganymede, allowing the two to be well decorrelated in the estimation. For the flybys, this is not the case. To mitigate this, we exploit the fact that the flybys are very short compared to the satellites' orbital periods. As a result, the influence of the uncertainties in the moons' positions when estimating the dynamics of JUICE during flybys is essentially that they are biased. This effect is included into our model by estimating per-arc biases for each of the data types.

For future work in which the \emph{absolute} uncertainties are a primary goal, the issues of the joint uncertainty of JUICE, the satellites and Jupiter must be more rigorously addressed. Additionally, optimizing the observation schedules of the various tracking data to maximize the science return of the mission will require such a combined analysis. For the planning of the VLBI data, this will include considerations of obtaining tracking data away from the flybys, to allow the arcs of the spacecraft dynamics to be more accurately joined in the constrained multi-arc methodology.

\begin{table}[tb]\footnotesize
\caption{List of conditional (fixed moon ephemerides) JUICE spacecraft position uncertainties in radial (R) along-track (A) and cross-track (C) direction during the three phases of the mission (nominal case in bold).}
\centering
\scriptsize
\begin{tabular}{l l l l | l l l l | l l l l }
\hline
\hline
\multicolumn{4}{ c|}{GCO500 [m]} & \multicolumn{4}{ c|}{GCO/GEO5000 [m]} & \multicolumn{4}{c}{Flybys [m]}\\
Case & R & A & C & Case & R & A & C & Case & R & A & C\\
\hline
1 & 0.2 & 2   & 1&1 & 1 & 5  & 2   &1 & 1 & 10 & 10\\
2 & 0.5 & 2   & 1&2 & 2 & 5  & 2   &2 & 2 & 10 & 10\\
3 & 0.5 & 2   & 2&3 & 2 & 10& 2   &3 & 2 & 20 & 10\\
\bf{4} & \bf{0.5} & \bf{5}   & \bf{2}&\bf{4} & \bf{2} & \bf{10}& \bf{5}   &\bf{4} & \bf{2} & \bf{20} & \bf{20}\\
5 & 1.0 & 10 & 5&5 & 5 & 20&10 &5 & 5 & 50 & 50\\
\hline
\end{tabular}
\label{tab:estimatedParameterTable}
\end{table}

\subsection{Data quality and weights}
\label{sec:dataWeight}
In this section, we discuss the inherent quality of the data that are used as input to the covariance analyses, as well as the associated weights used in the covariance analysis, which includes the JUICE orbit determination uncertainty (Section \ref{sec:juiceDynamics}). %We pay special attention to the manner in which we include the uncertainty of the spacecraft orbit uncertainty w.r.t. the moons, as we do not include the estimation of JUICE's orbit in-the-loop (Section \ref{sec:trackingData}).

Doppler data uncertainty for JUICE (denoted $\sigma_{\dot{r}}$) is expected to be at the level of 10 $\mu$m/s at an integration time of 60 s. The range measurements $r_{sc}$ obtained by 3GM are expected to have a measurement uncertainty (denoted $\sigma_{r}$) of 0.2 m \citep{Iess2013}. However, the measurement uncertainty of the range data is not purely white and uncorrelated. Thus, it can result in overly optimistic simulation results from a covariance analysis. An \emph{independent} range measurement can be created once every 5 minutes. 

The present uncertainty in VLBI measurements $\alpha$ and $\delta$ (collectively denoted $\sigma_{h}$) of planetary targets is at the 1.0 nrad($\approx$ 200 $\mu$as) level \citep{LanyiEtAl2007,DuevEtAl2012,JonesEtAl2015,ParkEtAl2015b,DuevEtAl2016}. However, current systems use an X-band signal, while for JUICE, both a Ka- and an X-band signal will be available. The use of the Ka-band signal, which has a wavelength that is nearly 4 times shorter than at X-band, could in principle result in observables that are 3-4 times more precise \citep{LanyiEtAl2007,CurkendallBorder2013}. We note that the use of K/Ka-band signals in VLBI is at present relatively uncommon, and sufficiently strong reference sources are sparser than at X-band \citep{MajidEtAl2008}. Moreover, the beam size scales linearly with wavelength resulting in a Ka-beam being ~4 times smaller than that of the same telescope at X-band.  Nevertheless, work on extending the celestial reference frame, operations and station capabilities, to K/Ka-band frequencies is ongoing \citep{HoriuchiEtAl2013,TremouEtAl2015,MalkinEtAl2015}. 

An additional promising technique for improving the precision of the VLBI observable is the use of in-beam phase referencing \citep[\textit{e.g.}][]{FomalontEtAl1999}. By using such a technique, the reference source and spacecraft signal are simultaneously observed by the telescopes, allowing improved data quality. However, the use of in-beam phase referencing at Ka-band will prove to be very challenging, since there are significantly fewer compact reference sources at this higher frequency \citep{MajidEtAl2008}.

To assess the influence of these possible improvements in VLBI data quality, we will analyze the uncertainty in the ephemeris generation for measurement uncertainties of 0.1, 0.5 and 1.0 nrad. In doing so, we obtain a direct link between possible system improvements and the strengthening of JUICE's science return, providing a robust scientific case for whether or not the effort to improve the VLBI data quality of PRIDE-JUICE should be made.

Based on  the discussion in Section \ref{sec:juiceDynamics}, we synthesize direct Doppler/range/VLBI measurements of the Galilean moons (see Fig. \ref{fig:observableGeometry}) by incorporating the conditional uncertainty in $\mathbf{r}_{sc}^{i}$, denoted $\boldsymbol{\sigma}_{\mathbf{r}_{sc}^{i}}$, and uncertainty  in $\dot{\mathbf{r}}_{sc}^{i}$, denoted $\boldsymbol{\sigma}_{\dot{\mathbf{r}}_{sc}^{i}}$. We denote the synthetic direct measurements of range, Doppler, declination and right ascension as $\tilde{r}_{i}$, $\tilde{\dot{r}}_{i}$, $\tilde{\delta}_{i}$ and $\tilde{\alpha_{i}}$, respectively. Since $\boldsymbol{\sigma}_{\mathbf{r}_{sc}}\ll \boldsymbol{\sigma}_{\mathbf{r}_{sc}^{i}}$, we can simply add the uncertainty in the spacecraft's moon-centered position and velocity to the observation uncertainty, by projecting $\sigma_{\mathbf{r}_{sc}^{i}}$ onto range, right ascension and declination vectors, respectively. As a result:
\begingroup
\allowdisplaybreaks[4]
\begin{align}
\sigma_{\tilde{r}}&=\sigma_{r}+\boldsymbol{\sigma}_{\mathbf{r}_{sc}^{i}}\cdot \hat{\mathbf{r}}_{i}\label{eq:totalRangeUncertainty}\\
\sigma_{\tilde{\dot{r}}}&=\dot{\sigma}_{r}+\boldsymbol{\sigma}_{\dot{\mathbf{r}}_{sc}^{i}}\cdot \hat{\mathbf{r}}_{i}\label{eq:totalRangeRateUncertainty}\\
\sigma_{\tilde{\alpha}}&=\sigma_{\alpha}+\frac{\boldsymbol{\sigma}_{\mathbf{r}_{sc}^{i}}\cdot \hat{\mathbf{r}}_{\alpha}}{r_{i}}\label{eq:totalRaUncertainty}\\
\sigma_{\tilde{\delta}}&=\sigma_{\delta}+\frac{\boldsymbol{\sigma}_{\mathbf{r}_{sc}^{i}}\cdot \hat{\mathbf{r}}_{\delta}}{r_{i}}\label{eq:totalDecUncertainty}
\end{align}
\endgroup
We stress that the JUICE orbit will be determined arc-wise in the actual data analysis, whereas we determine the dynamics of the moons over a single arc. As noted in Section \ref{sec:juiceDynamics}, the fact that the marginal and conditional uncertainties of both the satellites and spacecraft dynamics are not equal, especially for the flybys, is (to first order) mitigated by including per-arc estimation of observation biases.

%AAAA insert discussion of chaotic orbit determination AAAAAA.
%Such an approach is analogous to that used by \cite{DuxburyCallahan1988,LaineyEtAl2007} for processing optical astrometry data from spacecraft in the generation of Martian satellite ephemerides. 

We show the values we used for the components of $\boldsymbol{\sigma}_{\mathbf{r}_{sc}^{i}}$ and  in Table \ref{tab:estimatedParameterTable}. We distinguish between orbit determination during the flybys, GCO/GEO5000 phase and the GCO500 phase. Furthermore, we set different uncertainties on the radial, along-track and across-track components. These values are assumed constant per simulation case/mission phase. Clearly, the values provided in the table provide a broad range of uncertainties, which are in part derived from the performance of the Cassini mission during flybys \citep{AntreasianEtAl2008}, as well as numerical simulations (for the GCO500 phase) performed by \cite{ParisiEtAl2012}, as also discussed by \cite{DirkxEtAl2016}. The uncertainties in spacecraft velocities  $\boldsymbol{\sigma}_{\dot{\mathbf{r}}_{sc}^{i}}$ are obtained from $\boldsymbol{\sigma}_{\mathbf{r}_{sc}^{i}}$, scaled using the appropriate characteristic time $\Delta T$. In the case of the flybys, $\Delta T$ is the tracking arc duration. For the orbit phases, $\Delta T$ is the orbital period. For the flybys, the uncertainties in the along- and cross-track direction positions map directly to the corresponding velocities, as does the root-sum-square of the radial and along-track direction. The scaling factor for the flybys is $\Delta T/4$. For the orbit phase, the situation is the same, with the exception that the uncertainty in radial position maps to uncertainty in along-track velocity. The scaling factor for the orbit phase becomes $\Delta T/2$. In \ref{app:giacomo}, we compare our values in Table \ref{tab:estimatedParameterTable} to preliminary analysis of JUICE orbit determination. 

{The Earth-based astrometry data can be realized directly, or derived from photometry obtained during eclipses/ mutual events, which provides more accurate measurements of the relative satellite positions \citep{ArlotEtAl2014}. The photometric observations can only be obtained during periods of specific observational geometry, though. The recently proposed method of mutual approximations \citep{MorgadoEtAl2016,Emelyanov2017} may allow similar data to be obtained with a more relaxed observational schedule}. The direct astrometry from Earth will benefit from the Gaia star catalogue \citep{ArlotEtAl2012, RobertEtAl2017}, which should allow Earth-based absolute optical astrometric measurements of the Galilean moons at the 20 mas (100 nrad) level. In our simulations, we use 10, 20 and 50 mas measurement uncertainty for the Earth-based direct astrometry data. This corresponds to a linear position uncertainty of approximately 38 km, 75 km and 190 km at Jupiter, respectively (at a distance of 5.2 AU). 

The quality of the astrometric observations made by {the JUICE spacecraft}, however, are not limited by the quality of the star catalogue. The positioning of the Galilean moons from JUICE-based astrometry is complicated by the small distance at which the observations are made. As a result, the moons present themselves as extended bodies in the image plane, requiring a mapping from the center of figure to the center of mass of the moons \citep[\emph{e.g., }][]{PasewaldtEtAl2012}. This mapping will limit the accuracy to which the dynamics of the centers of mass of the moons can be measured. We use a linear measurement uncertainty of 5, 10 and 20 km for $\alpha_{i}^{sc}$ and $\delta_{i}^{sc}$.

\section{Ephemeris uncertainty simulation}
\label{sec:ephemerisMethodology}	
Here, we describe the approach we take to simulate the uncertainty of the ephemerides of the Jovian system from JUICE tracking data.  In Section \ref{sec:estimatedParameters}, we discuss the parameters that we consider in our simulations. We make use of covariance analysis, the details for which are discussed in Section \ref{sec:covarianceAnalysis}. The observation scheduling that we use in our simulations is discussed in Sections \ref{sec:obsPlanning}. Finally, we discuss the inherent rank deficiency in our estimation problem, as well as mitigation strategies, in Section \ref{sec:aprioriInformation}. We note that the long-term improvement in satellite ephemerides depends not only on the results of the JUICE mission, but also on the tracking data and estimation quality of related missions such as Juno, Europa Clipper, \emph{etc.}, as well as ground-based campaigns both before and after the JUICE mission. As such, realistically propagating the ephemeris uncertainties over long time periods is well beyond the scope of this work, as it would require the analysis of a range of additional existing and upcoming data sets.

\subsection{Estimated parameters}
\label{sec:estimatedParameters}

A detailed analysis of the sensitivity of the moons' dynamics to various perturbations over the timeframe of the JUICE mission was performed by \cite{DirkxEtAl2016}. Based on their analysis, we include the $k_{2}/Q$ of Io, Jupiter and Europa in the estimation (each at their respective once-per-orbit frequency). Additionally, we include Io's $k_{2}$ in our estimation. We estimate the initial state of the four Galilean moons in a Jupiter-centered frame, as well as the initial state of Jupiter in a barycentric frame. %Although it was found by \cite{DirkxEtAl2016} that the \emph{a priori} error in Jupiter's ephemeris will not impact the dynamics of the moons, the estimation of Jupiter's state is included in our analysis, as most of our observables (with the exception of JANUS astrometry) are influenced directly by both Jupiter's barycentric dynamics, in addition to the Galilean moons' local dynamics, see Eqs. (\ref{eq:measurementComposition}) and (\ref{eq:measurementComposition2}). %We provide some details on the mathematical formulation of the associated variational equations that are used for the ephemeris generation in \ref{app:varEq}.
%We propagate the dynamics of the moons and Jupiter using the model described by \cite{DirkxEtAl2016}, which is based on the model by \cite{LaineyEtAl2004}. The state transition and sensitivity matrices are propagated using the models described by \cite{MontenbruckGill2000}. We calculate all parital derivatives analytically. The formulation for the mixed problem of estimating both the barycentric position of Jupiter, and the Jupiter-centered position of the moons, is presented in Appendix \ref{sec:mixedEstimation}.
To mitigate the difference between the marginal and conditional uncertainties (Section \ref{sec:juiceDynamics}), as well as to absorb systematic errors in the measurements themselves, we estimate an arcwise observation bias for each of the data types. %By assessing the influence of the bias estimation, the contribution of systematic measurement and modelling (of the spacecraft state) errors to the estimation quality is assessed, although at a first-order level. 
%By including the bias estimation, we implicitly assume that the non-Gaussianities in the measurement errors can be modelled as an arcwise systematic error.

\subsection{Covariance analysis}
\label{sec:covarianceAnalysis}
%To estimate the uncertainty of the estimated states of the Galilean moons and Jupiter, as well as the associated physical parameters (Section \ref{sec:estimatedParameters}), 
We use simulated observations of the types presented in Section \ref{sec:dataModelling} as input to a covariance analysis \citep[\textit{e.g. }][]{MontenbruckGill2000,MilaniGronchi2010}.  Some mathematical details of covariance analysis are discussed in \ref{app:lsqOd}. %The weights and planning of the observations are presented in Sections \ref{sec:dataWeight} and \ref{sec:obsPlanning}, respectively. 
We use the dynamical model presented by \cite{DirkxEtAl2016}, which derives strongly from \cite{LaineyEtAl2004,LaineyEtAl2009}. We use a modified version of the Tudat toolkit\footnote{http://github.com/tudat}, discussed in more detail by \cite{Dirkx2015}.

For our analysis, we restrict ourselves to covariance analyses. Although the formal errors  that are obtained from such analyses are known to often be too optimistic, covariance analysis is well suited to our investigation, since our goal here is first and foremost to analyze the {relative} contribution of the VLBI data, not to obtain absolute values for the positioning uncertainties. Secondly, the broad range of settings that we use requires the analysis of many ($>$100,000) different scenarios. This would be infeasible with methods in which the true error is more realistically assessed, such as those used by \cite{KonoplivEtAl2011,LemoineEtAl2014,FiengaEtAl2014,DirkxEtAl2015}.

The ratio between the true and formal errors is difficult to robustly quantify for a general problem. It is estimated by \cite{JonesEtAl2015} (in the context of VLBI data analysis) that planetary ephemerides typically have a true-to-formal error ratio of 2-3. \cite{HofmannEtAl2010} estimate a value of 2 for the true-to-formal error ratio for estimation of relativity parameters using Lunar Laser Ranging (LLR) data. Conversely, \cite{LaineyEtAl2009} find that the formal error that they obtain for Io and Jupiter's $k_{2}/Q$ is a robust measure for the true uncertainty in this parameter. \cite{FiengaEtAl2015} find that the $3\sigma$ formal confidence interval obtained from a least squares adjustment corresponds well to the estimated true $1\sigma$ error, for general relativity parameters estimated from planetary ephemerides, provided that sources of model uncertainty are concurrently considered: specifically asteroid masses and observational biases.  

\begin{table*}[tb]\scriptsize
\caption{List of observation settings that are varied in the simulations (nominal case in bold).}
\centering
\begin{tabular}{l l l l }
\hline
\hline
VLBI cadence (Ganymede phase) & VLBI cadence (flybys) & {JUICE-based} astrometry & Bias estimation\\
\hline\\
Once per week & Every flybys & Io only, $\sigma=5$ km & \bf{Once per pass, all data types}\\
\bf{Once per month }& \bf{Every $2^{nd}$ flyby (per moon) }& Io only, $\sigma=10$ km & Once per pass, VLBI only\\
Once per 3 months & Every $3^{rd}$ flyby (per moon)  & Io only, $\sigma=20$ km & Once per pass, all data types except VLBI\\
None & Every $2^{nd}$ flyby/None (Callisto)  & Io and Europa, $\sigma=5$ km & None\\
 & Every $2^{nd}$ flyby/None (Europa)  & \bf{Io and Europa, $\sigma=10$ km }& \\
 & Every $2^{nd}$ flyby/None (Ganymede)  & Io and Europa, $\sigma=20$ km& \\
\hline
\hline
\end{tabular}
\label{tab:parameterSettingsTable}
\end{table*}

The use of covariance analysis in preliminary mission tracking performance has been conducted for various previous, upcoming and planned missions \citep[\textit{e.g.}][]{WuEtAl2001,RosatEtAl2008,DirkxEtAl2014,MazaricoEtAl2015}, with a focus on a variety of different data types. However, such studies have typically been on the direct science return of orbiter/flyby spacecraft tracking (gravity fields, rotational parameters \emph{etc}.), whereas our focus is on the analysis of natural satellite dynamics and the associated physical parameters of the Jovian system. Covariance analysis provides more realistic results for the analysis of the estimation of natural bodies (compared to orbiter dynamics), as the dynamical models that we use for the moons will be able to capture the full observable behaviour, as analyzed by \cite{DirkxEtAl2016}. For orbiter dynamics simulation and estimation, on the other hand, the dynamical model errors are not captured by a covariance analysis. Indeed, true-to-formal error ratios of parameters estimated directly from spacecraft tracking (gravity fields, rotational parameters \emph{etc,}) are typically in the order of 10 \citep[\emph{e.g.}][]{MartyEtAl2009,KonoplivEtAl2011,MazaricoEtAl2015}.

%Although we can model the dynamics of the moons sufficiently well for the covariance analysis, this is not necessarily the case for the measurement uncertainty (Section \ref{sec:dataWeight}). The use of covariance analysis assumes that all measurements are identically and independently distributed. In the case where the influence of the spacecraft's orbital uncertainty $\sigma_{\mathbf{r}_{sc}^{i}}$ is strong, see Eqs. (\ref{eq:totalRangeUncertainty}-\ref{eq:totalDecUncertainty}), this assumption will lead to overly optimistic results. This is especially true for highly accurate observations, specifically the range measurements and the VLBI measurements at 0.1 nrad precision. Consequently, the results we obtain here for the contribution of the VLBI data will be a conservative estimate, as the model errors that we neglect in our analysis will impact the range data more strongly than the VLBI data.

We compute the covariance for each of the combinations of the cases listed in Table \ref{tab:parameterSettingsTable} (and discussed in Section \ref{sec:obsPlanning}), as well as each combination of the JUICE position and velocity error cases shown in Table \ref{tab:estimatedParameterTable}. For each of these combinations, we simulate a solution where Doppler/range observations are the only radio data types, as well as a solution in which the VLBI data is included, at $\sigma_{h}$=0.1, 0.5 and 1.0 nrad.

\subsection{Observation planning}

\label{sec:obsPlanning}
%Although the definition of the JUICE mission scenario is far from final, the overall mission architecture is at present clearly defined (Section \ref{sec:juicePlanning}).

%Both the 3GM and PRIDE experiments are reliant on the reception of a radio signal transmitted by the spacecraft's MGA or HGA. The use of the MGA allows simultaneous operation of remote sensing instruments during flybys. This is not possible with the HGA, as its orientation w.r.t. the spacecraft is fixed. 

%Our focus in this work is to analyze the influence of the PRIDE-JUICE lateral position observables on the quality of the ephemerides of the Jovian system, and the associated physical parameters that will be generated from JUICE data. Consequently, 
We fix the 3GM range and Doppler data schedule and quality (see Section \ref{sec:dataWeight}) to a single nominal scenario:
\begin{itemize}
\item Flyby phase: an 8 hour Doppler and range tracking pass for each flyby, centered at closest approach: $\sigma_{\dot{r}}=0.01$ mm/s at 60 s integration time, and  $\sigma_{r}=0.2$ m, every 300 s.
\item Orbit phase: one 8 hour  Doppler and range tracking arc per day: $\sigma_{\dot{r}}=0.01$ mm/s at 60 s integration time, and  $\sigma_{r}=0.2$ m every 300s.
\end{itemize}
%The 300 s integration time is required to obtain an \emph{independent} measurement.

%By varying the planning and quality of the PRIDE lateral position data, we gain key insight into the quantitative influence of these data. Moreover, we provide a starting point from which to maximize PRIDE-JUICE's science return, by providing input into the further definition of the PRIDE observation scheduling, as well as quantifying the consequences of ground segment improvements. 

Due to the greater amount of resources that are required for PRIDE operations (many telescopes operating in synchronization) compared to 3GM, the cadence of the VLBI observations is substantially sparser than that of the Doppler and range observations. Nominally, we simulate VLBI tracking observations during every $2^{nd}$ flyby of each moon, and during a single 8 hour arc per month during the entire Ganymede orbit phase. We use the same tracking pass length and integration time as for the range measurements. %As discussed in Section \ref{sec:dataWeight}, we use lateral position uncertainties  $\sigma_{h}$ of 0.1, 0.5, 1.0 nrad.

In the first two columns of Table \ref{tab:parameterSettingsTable}, we summarize the variations of the PRIDE observation planning schedule that we investigate. We distinguish between the orbit and flyby phase, changing the cadence of the tracking observations during the orbit from weekly to none at all. For the flyby tracking, we investigate both an overall reduction/increase in tracking cadence, as well as the complete removal of VLBI tracking data from each of the moons. %The results from this analysis will provide insight into the value of the VLBI data during the varioius mission phases/flybys.

Although our focus in this work is the contribution of PRIDE-JUICE, we also investigate the influence of the use of optical astrometry. For the {JUICE-based} astrometry data, we simulate two arcs of data per year, either for Io only or Io and Europa (see Table \ref{tab:parameterSettingsTable}). For the simulated Earth-based astrometry, we apply a similar twice-per-year observation schedule.

%For all observations, the simulated measurements are constrained by occultations by each of the Galilean moons and Jupiter, causing occasional link outage. Also, we include a Sun avoidance angle constraint of AAA.

\subsection{Stabilization of Solution}
\label{sec:aprioriInformation}
The JUICE tracking data is obtained mostly at Gany\-mede, during the orbit phase. This provides a severe challenge for the estimation, since the dynamics of Io, Europa and Ganymede have to be estimated largely from observing the dynamics of Ganymede. The coupling between Callisto's dynamics and that of the inner three moons is rather weak, so the Callisto flyby data is not expected to contribute strongly to the estimation of the inner moons. %Conversely, the Callisto flyby data is well suited to estimating the dynamics of Callisto itself. 
The two flybys of Europa will be helpful in stabilizing the solution for the inner three moons, but the fact they are so closely spaced in time  reduces their contribution to constraining the dynamical evolution. In fact,  we find that the numerical inversion becomes strongly ill-posed (\emph{e.g.} normal equations have a condition number that approaches and goes beyond $10^{16}$), especially for the simulations without any optical data\footnote{Note that in our inversion, we scale the parameters to ensure that their partial derivatives are in the range [-1,1]. Consequently, the remaining poor conditioning is due to the physical signatures of the estimated parameters.}. %Therefore, special care needs to be taken to ensure the stability of the solution.

%No VLBI data & 8.5&44.2&178.9&98.0&173.3&63.9&38.4&467.6&47.3&1267.5\\
%With VLBI data $\sigma_{h}$ = 1.0 nrad & 8.5&34.9&181.8&98.0&177.9&62.3&19.3&118.3&25.0&73.3\\
%With VLBI data $\sigma_{h}$ = 0.5 nrad & 8.3&24.9&180.6&98.0&182.1&60.1&18.7&66.5&24.2&44.0\\
%With VLBI data $\sigma_{h}$ = 0.1 nrad & 7.9&6.6&169.4&97.8&112.3&35.1&17.2&19.0&18.1&16.5\\\hline
%\hline
%

%Firstly, to stabilize the solution we include optical astrometry data of both Io and Europa, from both the JA\-NUS instrument and Earth-based telescopes (Section \ref{sec:trackingData}). These observations provide direct measurements of the states of Io and Europa (although with much lower linear accuracy than the radio data), and they will be invaluable in decorrelating the dynamics of the inner three moons in the estimation. We discuss the influence of these data on the solution in Section \ref{sec:astrometryContribution}. However, we find in our simulations that \emph{only} including optical astrometry still often leads to an unstable estimation problem. Therefore, additional measures are required to stabilize the inversion.

As a first mitigation strategy, we make use of \emph{a priori} covariance for the estimated parameters. The present (and expected near-term) uncertainty in the dynamics of the Galilean moons is in the order of 10-100 km, obtained largely from optical astrometry observations \citep{LaineyEtAl2009}. However, applying these levels of \emph{a priori} uncertainty to the estimation will have relatively little influence on the solution, as the measurements are weighted at 2-5 orders of magnitude smaller linear uncertainty, making the influence of the  \emph{a priori} information small. 

However, the combination of the range and VLBI measurements from 3GM and PRIDE provide a direct three-dimensional position of the JUICE spacecraft w.r.t the Earth. By combining this with the JUICE spacecraft orbital solution w.r.t. the moon under consideration (Section \ref{sec:dataWeight}), we kinematically obtain a measurement of $\mathbf{r}^{I}_{0}+\mathbf{r}^{J}_{i}$ at a given measurement time, at an uncertainty of approximately that of the weights given by Eqs. (\ref{eq:totalRangeUncertainty})-(\ref{eq:totalDecUncertainty}). Using this knowledge of the kinematics, we can constrain the three-dimensional combined position of the moon and Jupiter at a given epoch. Using this approach, we apply an \emph{a priori} constraint for the position of both Europa and Jupiter at a level determined by the total range and VLBI weights, with an \emph{a priori} correlation of -1 between the initial state of Europa and Jupiter.

For the dissipation parameters of Io and Jupiter, we set the \emph{a priori} uncertainty at 3 times the formal error reported by \cite{LaineyEtAl2009}. We set the \emph{a priori} uncertainty of Europa's $k_{2}/Q$  at 0.075, which is a wide upper bound that is obtained from geophysical models, as calculated using the methods of \cite{JaraOrueEtAl2011}. The $k_{2}$ Love number of Io is observationally unconstrained. Models for the interior of Io predict Love numbers in a broad range of values (0.04-0.8). We conservatively set the \emph{a priori} uncertainty of Io's $k_{2}$ at 1.0. %For the measurement biases, we set an \emph{a priori} constraint equal to half the associated measurement uncertainty derived from Eqs. (\ref{eq:totalRangeUncertainty})-(\ref{eq:totalDecUncertainty}).

\begin{table*}[tb!]
\centering
\caption{Absolute values of formal initial position uncertainties in their orbital plane (IP) and perpendicular to their orbital plane (OP), for various values of $\sigma_{h}$ and $\sigma_{\mathbf{r}_{sc}^{i}}$. }
\begin{footnotesize}
\begin{tabular}{r|l l|l l|l l|l l|l l}
\hline
\hline
& \multicolumn{10}{c}{Initial position formal errors [m]}\\
\hline
& \multicolumn{2}{ c|}{Ganymede}& \multicolumn{2}{ c|}{Io}& \multicolumn{2}{ c|}{Europa}& \multicolumn{2}{ c|}{Callisto}& \multicolumn{2}{ c}{Jupiter}\\
Measurement case& IP & OP& IP & OP& IP & OP& IP & OP& IP & OP\\
\hline
JUICE Position Error Case 1 , No VLBI &17.1&44.2&491&616&332&126&38.2&453&46.4&1.25$\cdot 10^{3}$\\
With VLBI, $\sigma_{h}$=0.1 nrad &15.4&8.83&472&590&162&41.9&14.1&18.6&18&17.1\\
With VLBI, $\sigma_{h}$=0.5 nrad &16.4&26.6&485&612&321&103&15.3&66.8&24.4&47.4\\
With VLBI, $\sigma_{h}$=1.0 nrad &16.6&36.1&487&614&321&115&16.1&118&25.2&78.3\\\hline
JUICE Position Error Case 4 , No VLBI &25.1&47.8&596&618&391&213&73.5&857&72.9&1.93$\cdot 10^{3}$\\
With VLBI, $\sigma_{h}$=0.1 nrad &22.9&9.01&550&592&177&44.5&24.6&23.4&22&18.2\\
With VLBI, $\sigma_{h}$=0.5 nrad &24.4&27.2&582&613&335&145&28.9&71.8&29.3&48.9\\
With VLBI, $\sigma_{h}$=1.0 nrad &24.7&37.6&590&615&391&180&29.6&131&30.2&81.1\\\hline
JUICE Position Error Case 5 , No VLBI &43.9&201&740&619&595&367&188&2.23$\cdot 10^{3}$&220&6.09$\cdot 10^{3}$\\
With VLBI, $\sigma_{h}$=0.1 nrad &36&9.79&704&605&231&51.7&54.9&41.2&48.7&31.4\\
With VLBI, $\sigma_{h}$=0.5 nrad &41.8&33.3&735&617&495&182&69.7&91.6&85.4&79\\
With VLBI, $\sigma_{h}$=1.0 nrad &42.5&61.8&738&618&559&272&73&151&102&119\\\hline
\end{tabular}
\end{footnotesize}
\label{tab:absoluteNominalVlbiContribution}
\end{table*}

\begin{table*}[tb!]
\centering
\caption{Relative influence of VLBI data uncertainty on formal initial position uncertainties of Galilean moons and Jupiter in their orbital plane (IP) and perpendicular to their orbital plane (OP). The results shown are calculated w.r.t. the solution without VLBI for the same JUICE Position Error Case (only absolute changes $>5$\% shown). Note that a negative change represents an {improvement} in ephemeris quality.}
\begin{footnotesize}
\begin{tabular}{r|l l|l l|l l|l l|l l}
\hline
\hline
& \multicolumn{10}{c}{Relative initial position formal errors [\%]}\\
\hline
& \multicolumn{2}{ c|}{Ganymede}& \multicolumn{2}{ c|}{Io}& \multicolumn{2}{ c|}{Europa}& \multicolumn{2}{ c|}{Callisto}& \multicolumn{2}{ c}{Jupiter}\\
Measurement case& IP & OP& IP & OP& IP & OP& IP & OP& IP & OP\\
\hline
JUICE Position Error Case 1, $\sigma_{h}$=0.1 nrad &-10.1&-80&--&--&-51.3&-66.8&-63.2&-95.9&-61.2&-98.6\\
$\sigma_{h}$=0.5 nrad &--&-39.7&--&--&--&-18.2&-60&-85.2&-47.4&-96.2\\
$\sigma_{h}$=1.0 nrad &--&-18.3&--&--&--&-9.24&-57.9&-73.9&-45.6&-93.7\\\hline
JUICE Position Error Case 4, $\sigma_{h}$=0.1 nrad &-9.04&-81.2&-7.72&--&-54.6&-79.1&-66.6&-97.3&-69.8&-99.1\\
$\sigma_{h}$=0.5 nrad &--&-43.1&--&--&-14.1&-32.1&-60.7&-91.6&-59.8&-97.5\\
$\sigma_{h}$=1.0 nrad &--&-21.3&--&--&--&-15.4&-59.7&-84.7&-58.6&-95.8\\\hline
JUICE Position Error Case 5, $\sigma_{h}$=0.1 nrad &-18&-95.1&--&--&-61.1&-85.9&-70.9&-98.2&-77.9&-99.5\\
$\sigma_{h}$=0.5 nrad &--&-83.4&--&--&-16.8&-50.3&-63&-95.9&-61.3&-98.7\\
$\sigma_{h}$=1.0 nrad &--&-69.2&--&--&-5.95&-26&-61.2&-93.2&-53.9&-98\\\hline
\end{tabular}
\end{footnotesize}
\label{tab:nominalVlbiContribution}
\end{table*}

\section{Results - Satellite ephemerides uncertainty}
\label{sec:results}

Using the simulation methodology outlined in Sections \ref{sec:dataModelling} and \ref{sec:ephemerisMethodology}, the Galilean moon formal position uncertainty has been calculated for a broad range of mission and observation settings, using only simulated data obtained during the JUICE mission. The general results are presented in Section \ref{sec:nominalResults}, while the influence of specific simulation settings is discussed in Section \ref{sec:resultsObservationSettings}. %The results for the nominal case with and without VLBI data are shown in Tables \ref{tab:absoluteNominalVlbiContribution} and \ref{tab:nominalVlbiContribution}. 

%These results are discussed in detail in Section \ref{sec:nominalResults}, followed by a discussion on the influence of the mission parameters in Section \ref{sec:resultsObservationSettings}. In Section \ref{sec:resultsParameterEstimation}, we discuss the uncertainties in the estimation of the tidal parameters of the Jovian system. Finally, we briefly discuss the results we obtain for the uncertainty in Jupiter's initial state in Section \ref{sec:jupiterResults}. We will generally restrict ourselves to results obtained when varying a \emph{single} mission or system parameters. %However we will highlight other regions in the parameter space that we explore, for specific cases where interesting behaviour is observed.

\subsection{Nominal PRIDE contribution}
\label{sec:nominalResults}
Here, we present the results of the simulations using the nominal observation planning schedule (Table \ref{tab:parameterSettingsTable}), while varying the VLBI data uncertainty $\sigma_{h}$. %Furthermore, we investigate the general influence of the JUICE spacecraft position uncertainty $\boldsymbol{\sigma}_{\mathbf{r}_{sc}^{i}}$ on the relative contribution of the VLBI data (Section \ref{sec:dataWeight}). We investigate the influence of the JUICE position uncertainty in more detail in Section \ref{sec:juiceOrbitQuality}. 
The results of the simulations are summarized in Table \ref{tab:absoluteNominalVlbiContribution} (absolute formal uncertainty) and Table \ref{tab:nominalVlbiContribution}, which shows the relative change in formal errors when adding the VLBI data, compared to solutions without VLBI.
\subsubsection{Ganymede}

%For high-to-moderate quality JUICE orbits (Position Error Cases 1 and 4; see Table \ref{tab:estimatedParameterTable}) 

For $\sigma_{h}$=0.5 or 1.0 nrad, the contribution of the VLBI data to Ganymede's in-plane component (\emph{i.e.} the component that lies in the satellite's orbital plane) estimation is negligible (Table \ref{tab:nominalVlbiContribution}). The $\sigma_{h}$=0.1 nrad solution, however, consistently produces an uncertainty of Ganymede's in-plane position that is about 10-20 \% better than the solution without VLBI. For off-nominal operational scenarios, this range varies slightly, but consistently lies between 5-25\% (except for cases where the number of VLBI observations at Ganymede is greatly reduced). The greatest improved is obtained for JUICE Position Error Case 5. Although this indicates that the VLBI data helps to make the ephemerides quality somewhat less susceptible to the orbit determination quality of the spacecraft itself, the effect is only above our 5 \% cutoff at $\sigma_{h}=0.1$ nrad. Generally, at a maximum relative influence of 25 \% on the formal error, the contribution of the VLBI data on Ganymede's in-plane position is rather weak.

The uncertainty in Ganymede's out-of-plane component (\emph{i.e.} perpendicular to the satellite's orbital plane) is significantly reduced by the presence of the VLBI data, and there is a strong relation between the position uncertainty and $\sigma_{h}$. This is a result of the fact that the range and Doppler data contribute only weakly to this component, and that Ganymede's out-of-plane component correlates only weakly with other parameters. For the 1.0 nrad solution, increasing the uncertainty  in JUICE's position during the orbit phase from case 1 to case 5 degrades the solution by about 25 m, compared to a degradation of 150 m in the case of the no-VLBI solution. For the 0.1 nrad observable, the solution becomes almost independent of the orbit uncertainty during the orbit phase, and a formal position uncertainty at the 8-10 m level is consistently obtained. However, at these levels of uncertainty, the model simplifications that we have used may introduce more pronounced errors, in particular our approach in decoupling the orbit determination from the ephemeris generation.

\begin{table*}[tb!]
\centering
\caption{Relative influence of VLBI observation planning on initial position uncertainties of Galilean moons and Jupiter in their orbital plane (IP) and perpendicular to their orbital plane (OP), compared to nominal observation planning including VLBI at the given uncertainty (only absolute changes $>5$\% shown). Note that a negative change represents an {improvement} in ephemeris quality.}
\begin{footnotesize}
\begin{tabular}{r|l l|l l|l l|l l|l l}
\hline
\hline
& \multicolumn{10}{c}{Tracking Cadence}\\
\hline
& \multicolumn{2}{ c|}{Ganymede}& \multicolumn{2}{ c|}{Io}& \multicolumn{2}{ c|}{Europa}& \multicolumn{2}{ c|}{Callisto}& \multicolumn{2}{ c}{Jupiter}\\
Measurement case& IP & OP& IP & OP& IP & OP& IP & OP& IP & OP\\
\hline
Every flyby, $\sigma_{h}$=0.1 nrad &--&-13.6&--&--&--&-22.7&-9.6&-17.9&-11.9&-23.6\\
$\sigma_{h}$=0.5 nrad &--&-10.1&--&--&--&-9.02&--&-26.6&--&-21.3\\
$\sigma_{h}$=1.0 nrad &--&-5.52&--&--&--&--&--&-27.2&--&-20.8\\\hline
Every 3$^{rd}$ flyby, $\sigma_{h}$=0.1 nrad &--&--&--&--&--&--&--&--&--&--\\
$\sigma_{h}$=0.5 nrad &--&--&--&--&--&--&--&--&--&6.87\\
$\sigma_{h}$=1.0 nrad &--&--&--&--&--&--&--&--&--&8.98\\\hline
No Europa VLBI, $\sigma_{h}$=0.1 nrad &--&5.87&--&--&--&174&--&--&--&--\\
$\sigma_{h}$=0.5 nrad &--&--&--&--&--&15.8&--&--&--&--\\
$\sigma_{h}$=1.0 nrad &--&--&--&--&--&--&--&--&--&--\\\hline
No Ganymede flyby VLBI, $\sigma_{h}$=0.1 nrad &--&23.3&--&--&--&11.9&--&66.1&10.6&160\\
$\sigma_{h}$=0.5 nrad &--&9.71&--&--&--&--&--&29.1&--&146\\
$\sigma_{h}$=1.0 nrad &--&--&--&--&--&--&--&8.85&--&103\\\hline
No Callisto flyby VLBI, $\sigma_{h}$=0.1 nrad &--&23&--&--&--&--&104&2.07e+03&10.5&5.91\\
$\sigma_{h}$=0.5 nrad &--&--&--&--&--&--&91.8&541&6.01&--\\
$\sigma_{h}$=1.0 nrad &--&--&--&--&--&--&86.5&262&--&--\\\hline
Weekly Ganymede Orbit VLBI, $\sigma_{h}$=0.1 nrad &--&-16.8&--&--&--&--&--&--&-7.29&-15.2\\
$\sigma_{h}$=0.5 nrad &--&-45.8&--&--&--&--&--&--&--&-18.1\\
$\sigma_{h}$=1.0 nrad &--&-37.1&--&--&--&--&--&--&--&-14.5\\\hline
Trimonthly Ganymede Orbit VLB, $\sigma_{h}$=0.1 nrad &--&16.1&--&--&--&--&--&--&--&10.1\\
$\sigma_{h}$=0.5 nrad &--&12&--&--&--&--&--&--&--&7.97\\
$\sigma_{h}$=1.0 nrad &--&5.42&--&--&--&--&--&--&--&5.28\\\hline
No Ganymede Orbit VLBI, $\sigma_{h}$=0.1 nrad &--&103&--&--&--&--&--&12.5&5.59&42.8\\
$\sigma_{h}$=0.5 nrad &--&45.7&--&--&--&--&--&--&--&29.6\\
$\sigma_{h}$=1.0 nrad &--&15.8&--&--&--&--&--&--&--&19.1\\\hline
No Ganymede VLBI, $\sigma_{h}$=0.1 nrad &--&396&--&--&--&54.3&--&273&22.7&403\\
$\sigma_{h}$=0.5 nrad &--&65.3&--&--&--&8.01&--&67.3&7.53&227\\
$\sigma_{h}$=1.0 nrad &--&20.8&--&--&--&--&--&16.2&8.67&161\\\hline
\end{tabular}
\end{footnotesize}
\label{tab:planningInfluenceResults}
\end{table*}

\subsubsection{Callisto}
\label{sec:nominalCallistoInfluence}
Out of the four moons, the VLBI data provide the strongest improvements for Callisto (Table \ref{tab:nominalVlbiContribution}). %This was to be expected, since the many flybys of Callisto will provide ample opportunities to measure the instantaneous out-of-plane position of this moon. %This is in contrast to the observations at Europa, where only two flybys (at very small temporal separation) are planned. For Ganymede, on the other hand, the highly accurate orbit determination of the spacecraft during the orbit phase means that the influence of the range observable is much higher than of the lateral position observable (Section \ref{sec:dataWeight}), which is not the case for Callisto. 
The in-plane component of Callisto is consistently determined at $>$50 \% better when using the VLBI data for the nominal 1.0 nrad case, increasing to 70 \% for the 0.1 nrad VLBI data at JUICE Position Error Case 5. We see a stronger influence of the VLBI data for the poor JUICE orbit determination quality, indicating that the lateral position data helps to make the solution less sensitive to spacecraft orbit errors.

The uncertainty in the out-of-plane component of Callisto is improved even more by the inclusion of the PRIDE data, by a factor of at least 4 and up to a factor 50 for the extreme case of low-quality (Case 5) JUICE orbit determination and $\sigma_{h}$ = 0.1 nrad. The fact that the {relative} VLBI contribution becomes especially high for the poor JUICE orbit determination reiterates the result that inclusion of the VLBI data partly mitigates the influence of a reduction in JUICE orbit determination quality. Since the out-of-plane position of Callisto is very weakly coupled to the in-plane dynamics of Callisto, or the dynamics of any of the other moons, the no-VLBI solution relies heavily on the weak kinematic influence of the out-of-plane component on the line-of-sight observables. The lateral position observable, however, directly measures this component, resulting in a solution that scales almost directly with the uncertainty of the VLBI observable.

\subsubsection{Europa and Io}
\label{sec:nominalEuropaIoInfluence}
Since Io is not observed directly at all, and Europa is only observed twice at very closely spaced points in time, the constraints on both Io's and Europa's dynamics rely strongly on their signatures on the dynamics of Ganymede (due to the Laplace resonance), as well as on the direct optical astrometry data. Due to the strong dependence on the astrometry data (Section \ref{sec:astrometryContribution}), their orbital solution is largely insensitive to the radio tracking data quality (Table \ref{tab:nominalVlbiContribution}).

The in-plane component of Europa is estimated at about 50 \% better when including the VLBI data at 0.1 nrad. This is in part due to the stronger \emph{a priori} uncertainty that can be applied in these cases (Section \ref{sec:aprioriInformation}), which improves the solution beyond that achieved by mainly {JUICE-based astrometry} data. The out-of-plane component shows a more substantial improvement, which remains significant for the full range of cases that we analyze. Unlike the solution for Ganymede and Callisto, the out-of-plane uncertainty is actually \emph{smaller} than that of the in-plane component. This is a direct result of the strong correlations with the position of both Io and Gany\-mede that are present for the in-plane, but not the out-of-plane, component. As a result, the addition of VLBI data of Europa has a much more significant impact on its out-of-plane component. %This correlation of the in-plane components of Io and Europa causes a weak influence ($<$10 \% improvement) of the $\sigma_{h}$=0.1 nrad VLBI data  on the in-plane component of Io, despite the absence of direct observations of Io's position. %The out-of-plane component of Io is only very weakly correlated with the dynamics of the other moons. Nevertheless, the solution is dominated by the JANUS data, and is mostly independent of tracking data accuracy.

A numerical issue occurs with the determination of the uncertainty of the dynamics of Io (and to a lesser degree Europa). This is a consequence of the remaining degree of ill-posedness of the solution, despite the  steps taken to reduce the problem (Section \ref{sec:aprioriInformation}). In particular, the ephemeris solutions in certain cases show a strong sensitivity to simulation settings in a way that is indicative of the onset of numerical instability. %In particular, some of the results shown in Table \ref{tab:planningInfluenceResults} provide uncertainty increases/decreases where the opposite behvaiour would be expected. 

This %onset of illposedness, despite the various measures we have discussed in Section \ref{sec:aprioriInformation}, 
will provide a challenge when analyzing the JUICE tracking data. Possibly, the weights of the tracking data will need to be reduced to below the inherent measurement/JUICE position uncertainty \citep[\emph{e.g.},][]{MurrowJacobson1988}, giving greater weight to the \emph{a priori} constraints. However, this would prevent the full information content in the tracking data from being exploited. 

We do not focus on detailed analysis strategies by which to reduce the condition number of the normal equations. %Nevertheless, we will highlight this issue in the remaining results where relevant.
Future analyses, preferably including all tracking data, should be performed to ensure that the full suite of JUICE tracking data can be optimally used to address the mission's science goals. Nevertheless, the results we obtain here strongly indicate that PRIDE data may not significantly contribute to the generation of stable ephemerides of Io, and to Europa's ephemeris in only a limited fashion.

\subsection{Influence of observation settings}
\label{sec:resultsObservationSettings}

%Tables \ref{tab:estimatedParameterTable} and \ref{tab:parameterSettingsTable} summarize the different mission and observation settings that we have varied in our simulations. 
In this section, we discuss the influence that varying the observation settings has on the formal estimation error, and  in particular the relative influence of the VLBI data. Sections \ref{sec:obsPlanningResults}, \ref{sec:astrometryContribution} and \ref{sec:juiceOrbitQuality} present the influence of the VLBI observation planning (both during flybys and the orbit phases), the use of optical astrometry and the quality of the Doppler-based JUICE orbit determination, respectively.

\subsubsection{Influence of observation planning}
\label{sec:obsPlanningResults}
The results of influence of the VLBI observation planning are shown in Table \ref{tab:planningInfluenceResults}, where the results are shown in which {only} a single component of the planning under consideration is changed (Table \ref{tab:parameterSettingsTable}). 

%\begin{table*}[tb!]
%\centering
%\caption{Relative influence of optical astrometry on initial position uncertainties in their orbital plane (IP) and perpendicular to their orbital plane (OP), compared to nominal observation planning including VLBI at the given uncertainty (only changes $>$5\% shown).}
%\centering
%\begin{footnotesize}
%\input{Figures/tableAstrometryRelativePositionUncertainty.tex}
%\end{footnotesize}
%\label{tab:opticalAstrometry}
%\end{table*}

The Ganymede in-plane component is only negligibly influenced by the observation cadence during the orbit phase, as a result of the highly accurate 3GM data and spacecraft orbit determination. The out-of-plane component is improved in the range of $15-45 \%$ when increasing the cadence to weekly, as opposed to the nominal monthly. %There is little difference between the different $\sigma_{h}$ cases, indicating that the observation accuracy at Ganymede itself no longer limits the uncertainty in its out-of-plane components, but that correlations with the states of the other moons dominate the error budget. However, at the level of uncertainty that is obtained ($<$5 m), more detailed analysis of the various (non-Gaussian) error sources may be required.

Reducing the amount of VLBI data during the orbit phase to trimonthly increases the error in Ganymede's out-of-plane component by only $5 \%$ for the $\sigma_{h}$ = 1.0 nrad case, compared to just over 15 \% for the 0.1 nrad case. When omitting the VLBI data entirely during the orbit phase, the influence is again especially strong for the $\sigma_{h}=0.1$ nrad case (degradation of close to 100 \%). 
%, since the solution reduces almost to the range-only solution. 
This indicates that an improvement in VLBI tracking data quality beyond the current 1.0 nrad level not only improves the nominal uncertainty of the ephemerides (Table \ref{tab:nominalVlbiContribution}), but it consequently also provides a greater incentive for a denser tracking schedule. %Although this result may sound paradoxical, as one may expect a \emph{reduced} tracking schedule to be acceptable for improved VLBI tracking data. However, improving the VLBI data allows the solution to rely more strongly on the lateral position information, concordantly increasing the sensitivity of the solution to the data volume. 

Interestingly, removing the VLBI data altogether during the Ganymede orbit phase reduces the quality of Callisto's out-of-plane component in the case of the 0.1 nrad observable quality. This is a result of the (weak) dynamical coupling between the moons. The degradation is especially strong for relatively poor orbit quality at Ganymede, and relatively good orbit quality during the flybys. %The coupling between Callisto's out-of-plane component and Gany\-mede tracking is discussed in Section \ref{sec:juiceOrbitQuality} in some more detail. 
The same effect is observed in the inverse case: removing the Callisto tracking altogether significantly affects the uncertainty in Ganymede's out-of-plane component for $\sigma_{h}=0.1$ nrad.

In general, the removal of the VLBI data during the Callisto flybys reduces the uncertainty in both its in- and out-of-plane component to (almost) the no-VLBI solution. The VLBI data during the Europa flybys significantly affects the solution mostly for the $\sigma_{h}=0.1$ nrad case (Table \ref{tab:planningInfluenceResults}), with a relatively weak influence for the $\sigma_{h}$=0.5 nrad case. %, which is reflected in the results shown in Table \ref{tab:planningInfluenceResults} for the nominal influence of VLBI on the initial position error of Europa. 
Similarly, we observe that none of the variations of the VLBI planning that we consider here significantly affect the estimation error of Io.

The overall cadence of the VLBI tracking during the flybys is only weakly influential on the estimation quality in most cases. In fact, the influence is below our $5\%$ threshold for each satellite when going from the nominal case (every $2^{nd}$ flyby) to a slightly reduced every $3^{rd}$ flyby  (\emph{e.g} one out of every three). Performing VLBI tracking during every flyby is more influential, but only moderately so. Again, it is especially influential for highly accurate data ($\sigma_{h}=0.1,0.5$ nrad), with the exception of Callisto out-of-plane component. 
%The coupling between the dynamics of Ganymede and Callisto is visible in the results in which either the Ganymede or Callisto VLBI data is omitted. In fact, removing the Ganymede flyby data influence the estimation of Callisto's ephemeris more than Ganymede's ephemeris. 
By far the strongest relative influence of the flyby data comes from the direct contribution of the Callisto flybys (as previously shown in Section \ref{sec:nominalResults}).

%\begin{table*}[tb!]
%\centering
%\caption{Relative influence of bias estimation on initial position uncertainties, compared to nominal observation planning including VLBI at the given uncertainty (only changes $>$5\% shown).}
%\begin{footnotesize}
%\input{Figures2/tableAstrometryRelativePositionUncertainty.tex}
%\end{footnotesize}
%\label{tab:biasEstimation}
%\end{table*}

\subsubsection{Contribution of optical astrometry}
\label{sec:astrometryContribution}
%The results for the nominal simulation scenario, in which only the planning of the optical astrometry and $\sigma_{h}$ are varied, is shown in Table \ref{tab:opticalAstrometry}. 
%
%We do not show 
As shown in Table \ref{tab:parameterSettingsTable}, we have used various settings for the optical astrometry obtained by the JUICE spacecraft. Here, we qualitatively discuss the influence of these data on the estimation results. The results without the optical astrometry consistently become numerically unstable. We find that when using only the 3GM and PRIDE data, condition numbers at the level of $10^{16}$-$10^{18}$ are commonplace in these cases. %, so the use of JANUS optical astrometry is crucial in stabilizing the solution. The physical origin of the ill-posedness is a result of the strong coupling of the dynamics of Io, Europa and Ganymede (Sections \ref{sec:aprioriInformation} and \ref{sec:nominalEuropaIoInfluence}). %Without optical astrometry, the dynamics of these inner three moons would have to be determined almost exclusively from spacecraft tracking at Ganymede (both during the orbit and the flybys). The Europa flybys will contribute to the solution, but the fact that they are very closely spaced in time strongly degrades their ability to fix the orbit of Europa.
Our results indicate that the astrometry data of especially Io is crucial in stabilizing the solution, while the Europa optical astrometry data is of secondary importance. In future work, the planning of the JANUS {and/or NavCam} astrometry as input to the ephemerides should be analyzed in more detail and optimized.

We have also performed simulations in which Earth-based astrometry is included, to assess the possible contribution of these data to the ephemeris generation. However, even for the case of 10 mas optical astrometry from Earth, the data often fails to stabilize the solution, since the linear position uncertainty ($\approx$ 40 km; see Section \ref{sec:dataWeight}) is orders of magnitude worse than the radio and {JUICE-based} astrometry data. 

{However, we stress that the Earth-based optical data will be of crucial importance outside of the time interval that JUICE is in the Jovian system, to provide a long-period data set for the positions of the moons. A detailed investigation of the combination of spacecraft data, and a long time-series of Earth-based optical data, is beyond the scope of this article and will be investigated in future work. }

%To consistently obtain a converged solution in the absence of optical astrometry data, the weights that we have used for the radio tracking types would have to be reduced, so that the \emph{a priori} orbit uncertainty would have a stronger influence on the solution, consequently stabilizing the inversion. However, such an approach would not allow the full potential of the tracking data to be exploited, and it is strongly recommended that the JANUS astrometry be included in the JUICE operational scenario.

As a potential complementary approach to the use of optical astrometry to stabilize the solution, the use of JUICE data combined with data from the Europa Clipper mission, previously called EMFM (Europa Multiple Flyby Mission), \citep{ToddEtAl2015,MazaricoEtAl2015} may result in a stabilized solution for the ephemerides of the moons, as it would provide direct tracking data of Europa, spaced in time much wider than is the case for JUICE. Similarly, Juno data, as well as any future tracking data that provides accurate information on the dynamics of Europa or Io, such as the Io Volcano Observer (IVO) mission \citep{McEwenEtAl2014} will be greatly beneficial to the analysis of JUICE tracking data. This indicates that multiple missions to the Jovian moons will enhance the total science return of each mission separately, by providing measurements of the dynamics that are more evenly distributed over the various moons.

\begin{figure*}[tb!]
\centering
\subfigure[]{
\includegraphics[width=0.47\textwidth]{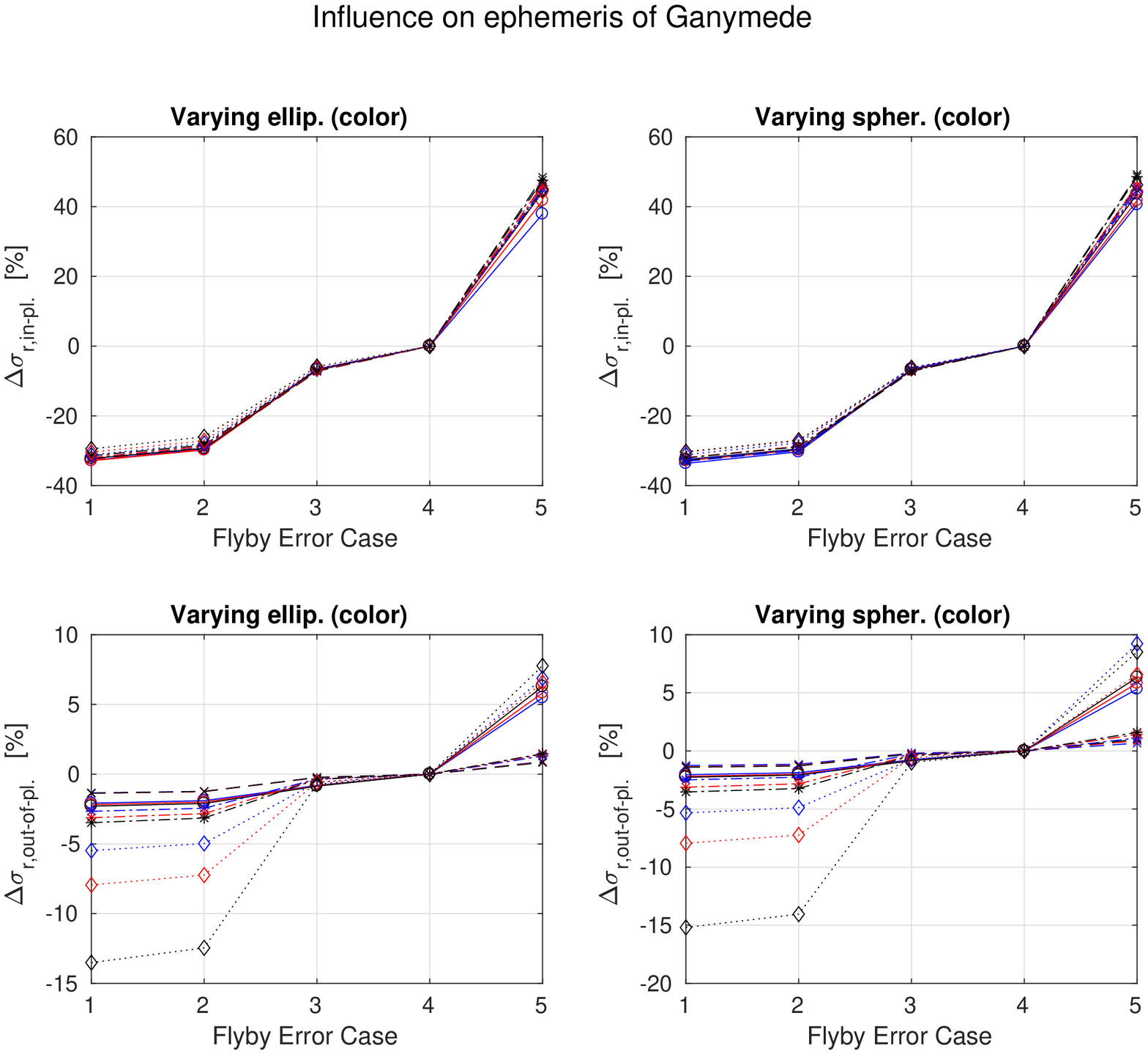}
\label{fig:juicePositionInfluenceGanymede}
}
\subfigure[]{
\includegraphics[width=0.47\textwidth]{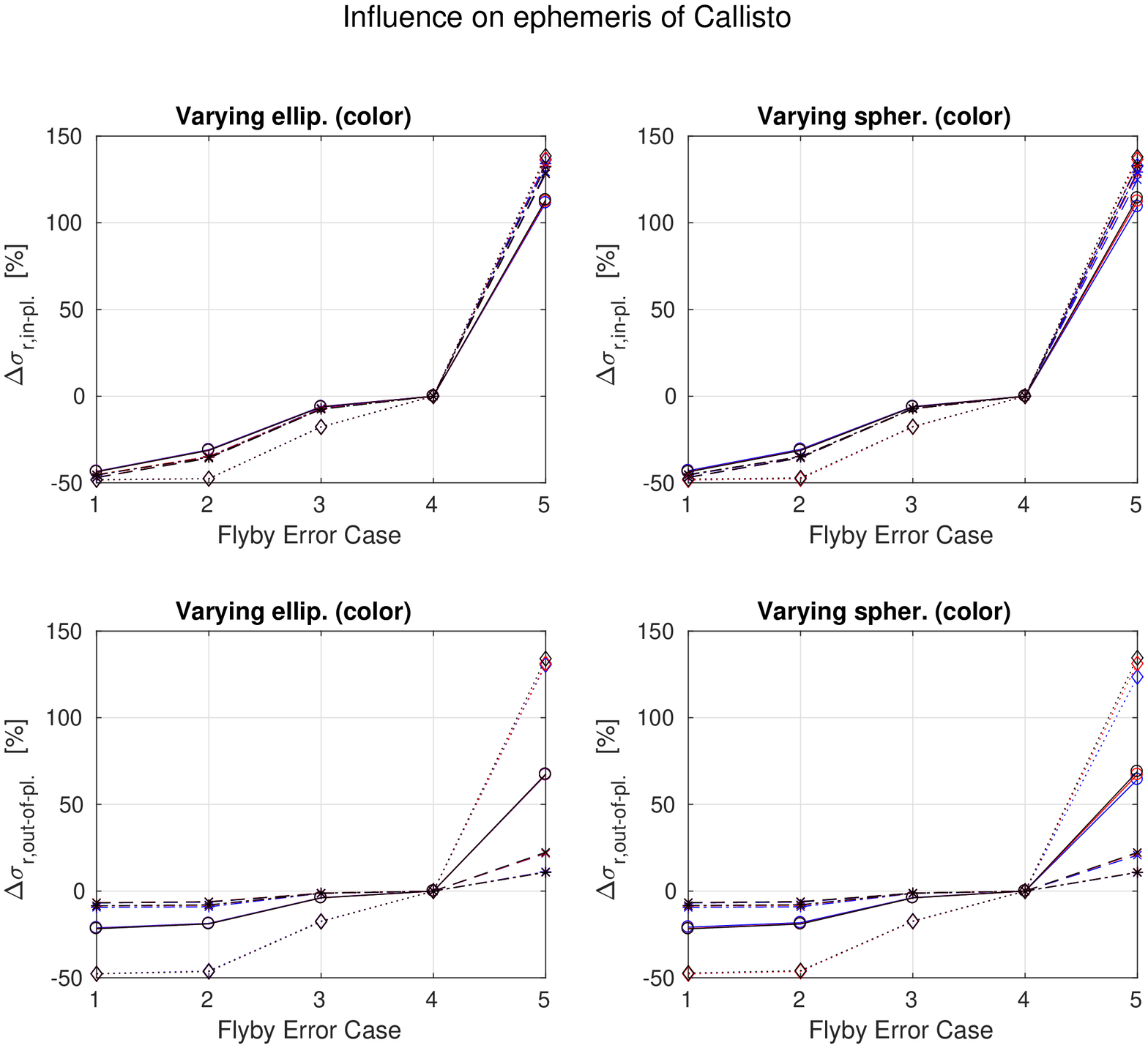}
\label{fig:juicePositionInfluenceCallisto}
}
\subfigure{
\includegraphics[width=0.23\textwidth]{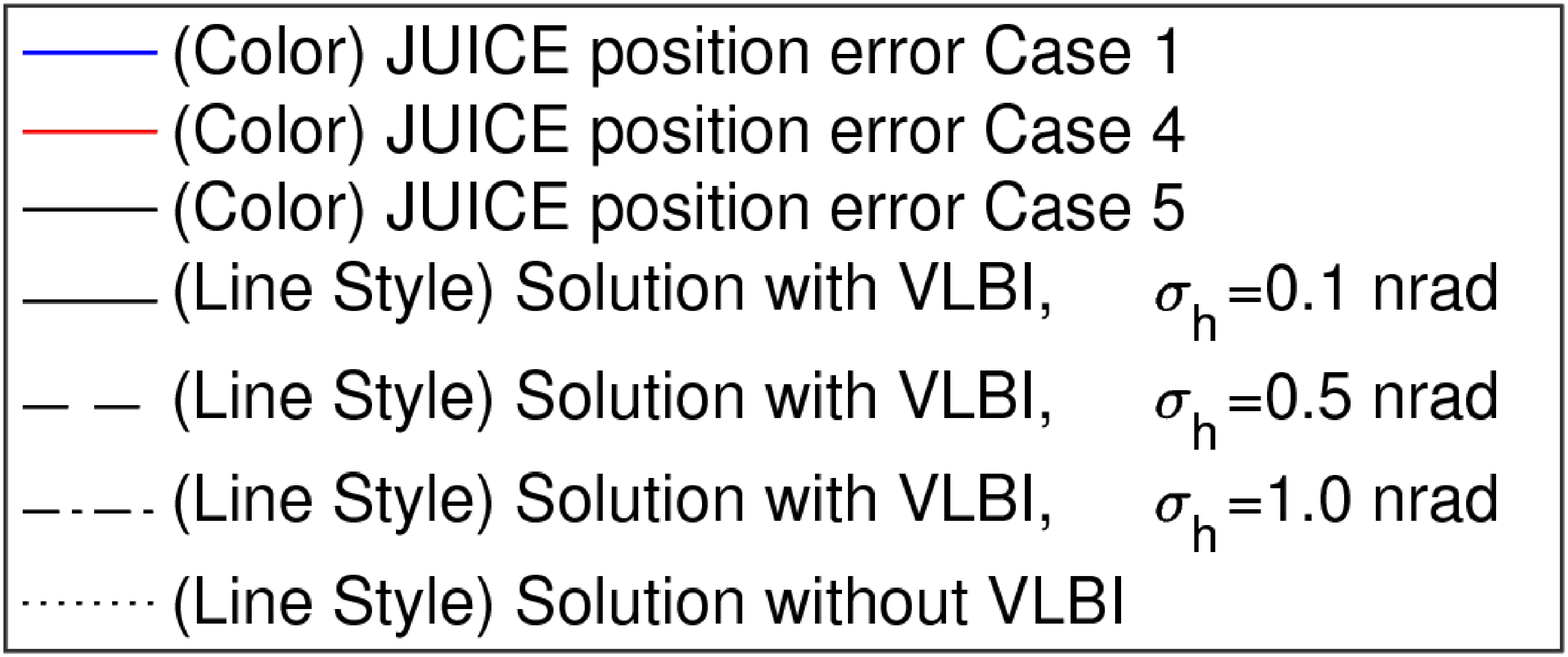}
}
\caption{Relative influence of JUICE orbiter positioning uncertainty during flybys on (a) Ganymede's and (b) Callisto's initial position estimation error. The positioning error cases are shown in Table \ref{tab:estimatedParameterTable}  (we take case 4 in each phase to be the nominal case). In each figure, one phase is varied along the abscissa, while Case 1, 4 and 5 of one other phase is shown in color (as indicated by figure caption). Settings for VLBI data are indicated by line style.}
\label{fig:juicePositionInfluence}
\end{figure*}

\begin{figure}[tbp!]
\centering
\includegraphics[width=0.47\textwidth]{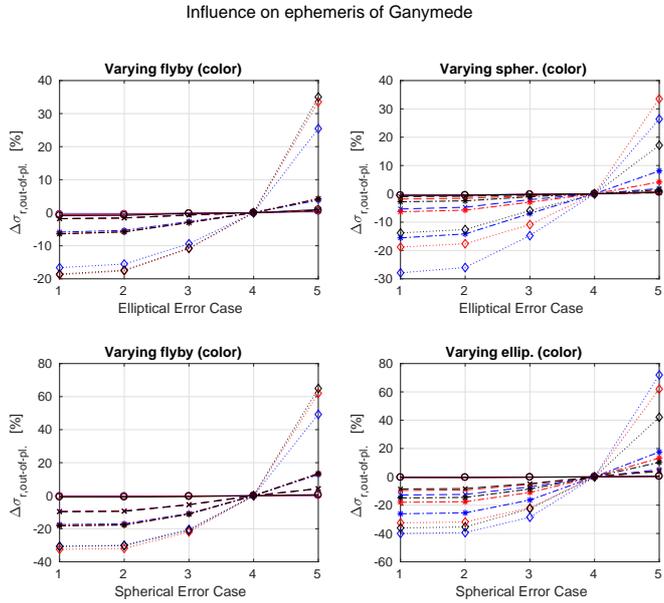}
\label{fig:juicePositionInfluenceGanymedeOp}
\caption{Relative influence of JUICE orbiter positioning uncertainty during elliptical and spherical orbit phase on estimates of out-of-plane initial position estimation error of Ganymede. The positioning error cases are shown in Table \ref{tab:estimatedParameterTable}, (we take case 4 in each phase to be the nominal case). In each figure, one phase is varied along the abscissa, while Case 1, 4 and 5 of one other phase is shown in color (as indicated by figure caption).  Settings for VLBI data are indicated by line style. The legend is given in Fig. \ref{fig:juicePositionInfluence}. }
\label{fig:juicePositionInfluence2}
\end{figure}

%\begin{figure*}[tbp!]
%\centering
%\includegraphics[width=0.9\textwidth]{Figures/JuicePositionInfluenceCallistoWrtSingleComponent.eps}
%\caption{Influence of JUICE orbiter positioning uncertainty on Callisto initial position estimation error. The positioning error cases are shown in Table \ref{tab:estimatedParameterTable}, where we take case 4 in each phase to be the nominal case. The figures show each combination of two mission phases, keeping the third mission phase set to the nominal case.}
%\label{fig:juicePositionInfluenceCallisto}
%\end{figure*}

\subsubsection{Influence of JUICE orbit quality}
\label{sec:juiceOrbitQuality}
We stack the position uncertainty of JUICE with the measurement uncertainty to obtain the measurement weights for the least-squares adjustment, see Eqs. (\ref{eq:totalRangeUncertainty})-(\ref{eq:totalDecUncertainty}). Here, we discuss the influence of the components of the position uncertainty on the moon initial position uncertainty (see Table \ref{tab:estimatedParameterTable}) in more detail.

We do not show all combinations of uncertainty cases explicitly, but limit ourselves to the those where the influence of the initial position uncertainty is consistently outside the $\pm 10$\% range. We have considered three different mission phases (flybys, high-altitude orbit and low-altitude orbit), with distinct uncertainties (Table \ref{tab:estimatedParameterTable}). %Figs.  \ref{fig:juicePositionInfluence} and  \ref{fig:juicePositionInfluence2} show the relative influence of changing the uncertainty in a single mission phase. 

The influence of the flyby positioning uncertainty on the ephe\-me\-rides of Ganymede and Callisto (in- and out-of-plane) is shown in Figs. \ref{fig:juicePositionInfluence}. The influence of the positioning error during the orbit phase on the out-of-plane component of Ganymede's ephemeris is shown in Fig. \ref{fig:juicePositionInfluence2}.The figures show the results for a number of combinations of uncertainties in other phases, as well as VLBI data uncertainty.  We omit the results for Io and Europa here entirely, as the influence of the radio tracking data is very limited for these bodies (Section \ref{sec:nominalEuropaIoInfluence}). %The figures show the relative change in moon initial position formal error, as a function of the cases shown in Table \ref{tab:estimatedParameterTable}. 

Figure \ref{fig:juicePositionInfluenceGanymede} shows that the flyby positioning uncertainty has a substantial influence on the quality of Gany\-mede's ephemeris, especially for the in-plane components. The influence on the in-plane components is only very weakly influenced by the quality of the VLBI data, however. For the out-of-plane component, the total variation with JUICE orbit quality is limited to about $\pm 10$\%, even in the absence of VLBI data. Nevertheless, the inclusion of VLBI largely removes the (small) dependency of the out-of-plane uncertainty on $\boldsymbol{\sigma}_{\mathbf{r}_{sc}^{i}}$. 

For Callisto, it is mostly the flyby tracking data that influence the solution (Fig. \ref{fig:juicePositionInfluenceCallisto}). The in-plane component varies with flyby positioning uncertainty in essentially the same manner for both the solutions without and with VLBI data. This magnitude of the influence of JUICE positioning uncertainty is substantially stronger for Callisto than for Ganymede. This is a result of the relatively stronger influence of $\boldsymbol{\sigma}_{\mathbf{r}_{sc}^{i}}$ and $\boldsymbol{\sigma}_{\dot{\mathbf{r}}_{sc}^{i}}$ on the measurement weight during the flybys (compared to the orbit phase). For the out-of-plane component, however, the inclusion of the VLBI data again strongly decreases the sensitivity of the positioning uncertainty, as was the case for Ganymede.

The influence of the JUICE positioning errors during the Ganymede phase are of relatively minor influence on the ephemerides of all moons except Ganymede itself. For the uncertainty in Ganymede's ephemeris, it is especially the out-of-plane component that is affected by JUICE's position error during the orbit phases (see Fig. \ref{fig:juicePositionInfluence2}), with changes of down to $-50\%$ and up to $+70\%$ in the absence of VLBI data, when moving from spherical orbit phase error case 1 to case 5.
%However, we see in AAA that the inclusion of VLBI data makes the out-of-plane solution much less susceptible to the quality of the orbit of the JUICE spacecraft, when compared to the solution without VLBI. 
The influence of spacecraft position uncertainty during the orbit phase on the out-of-plane component reduces to about 10 \% and 30 \% for the $\sigma_{h}$=0.1 nrad  and 1.0 nrad cases, respectively. %When estimating the out-of-plane dynamics from the range data alone, minor errors in the spacecraft orbit can be amplified greatly, due to the geometry of the estimation problem, explaining the distinct behaviour of the different situations. Also, the relative contribution of $\sigma_{\mathbf{r}_{sc}^{i}}$ to the range observable is much stronger than for the VLBI observable, even without the different geometric influence. 

By taking a broad range of values for the orbital uncertainties, we indirectly include the performance of the 3GM (and PRIDE) Doppler tracking into our simulation, albeit in a conceptual manner. Future detailed analyses of the JUICE orbit determination, including the various uncertainties in the environment (gravitational parameters, gravity field coefficients, radiation pressure \emph{etc.}), should be combined with our results here to analyze the coupling between the two in detail. Here, our focus is on the contribution of the VLBI data, not a full analysis of the tracking data inversion. In preparation of such detailed analyses, we contrast our assumptions and results with the preliminary results obtained from a covariance analysis of the spacecraft orbit determination from Doppler tracking data during the various mission phases using the \emph{Orbit14} software \citep{CicaloEtAl2016} (keeping the moons' states fixed) in \ref{app:giacomo}.

\section{Results - Jovian system parameter uncertainty}
\label{sec:jovianSystemResults}
In addition to satellite states, we also estimate the uncertainty of various physical properties of the bodies in the Jovian system, and Jupiter's barycentric dynamics, which we discuss in Sections \ref{sec:resultsParameterEstimation} and \ref{sec:jupiterResults}, respectively.

\subsection{Physical parameter estimation quality}
\label{sec:resultsParameterEstimation}
%As discussed in detail by \cite{DirkxEtAl2016}, the physical characteristics of the Jovian system that will be visible in the JUICE-derived ephemerides of the Galilean moons are limited. They conclude that only Io's $k_{2}/Q$ will be unambiguously visible, with possible observability of Jupiter's and Europa's $k_{2}/Q$. Furthermore, they show that there is a small possibility that Io's $k_{2}$ could be observable.

%Here, we present the results for the uncertainties of the physical parameters that we include in our simulations. 
The simulation results show that the dissipation inside Io can  be constrained at the level of $4.5\text{-}5\cdot 10^{-3}$ formal error by JUICE data, a value which is quite insensitive to most of our radio data observation settings. An important exception occurs when using lateral position data with a 0.1 nrad uncertainty, for which the uncertainty of Io's $k_{2}/Q$ decreases by about 20\% (for accurate JUICE orbit reconstruction) to $3.5\text{-}4.5\cdot 10^{-3}$.  %Also, we observe a weak, yet clear influence of the error in the GCO500 across track component of the spacecraft's state. The same effect was observed in the uncertainty of both Io's and Europa's initial state (Section \ref{sec:juiceOrbitQuality}). 
However, when omitting (or reducing the quality of) the optical astrometry data from the estimation, the quality and stability of the estimation of Io's $k_{2}/Q$ reduces to the \emph{a priori} uncertainty, indicating that the estimation is obtained largely from these data. This is directly in line with the results presented in Section \ref{sec:nominalEuropaIoInfluence}, where we show that the dynamics of Io (and to a lesser degree Europa) are only weakly influenced by the radio data, and are instead largely derived from the optical astrometry. %Even with our nominal astrometry observation scenario, the dissipation estimate shows some instability, as it is closely related to the estimation of Io's dynamics. %Clearly, an estimate of the dissipation in Io from JUICE data alone will improve upon the existing estimate produced by \cite{LaineyEtAl2009}

Moreover, considering the current uncertainty in the dissipation inside Io, combined with the expected improvement in astrometric data reduction facilitated by the Gaia mission \citep{ArlotEtAl2012}, it is unlikely that a JUICE-only solution will provide an improved estimate of the dissipation inside Io, compared to the expected state-of-the-art in the 2030s, especially when one considers that our results are only \emph{formal} errors. However, the JUICE data will provide a crucially accurate data set that will complement the existing and future optical astrometry data. Specifically, the JUICE tracking data will provide a 'ground truth' for the positions of the moons (especially Ganymede), providing the ephemeris generation with less freedom when fitting a long-period data set. Additionally, by optimizing the JANUS {and NavCam} observation schedule of Io, the dissipation estimation will likely be improved beyond the values we have obtained here. %Furthermore, the fact that JUICE tracking data is strongly biased towards the positioning of Ganymede means that the influence of Jupiter's dissipation is much less visible, and the estimates of $(k_{2}/Q)^{1,0}$ and $(k_{2}/Q)^{0,1}$ are only weakly correlated. This is a direct result of the fact that Io's dissipation causes a significant secular eccentricity rate in Io's orbit, which causes a much stronger change in Ganymede's orbit than the semi-major axis-only shoft caused by the dissipation inside Jupiter. 

Our simulations show that even when setting Europa's \emph{a priori} $k_{2}/Q$ to 0.075, the estimation fails to improve the uncertainty in a meaningful manner for any of the simulations. The same holds for the estimation of Io's $k_{2}$, which is not improved significantly beyond its \emph{a priori} value. Similarly, the dissipation inside Jupiter is only weakly improved beyond its \emph{a priori} uncertainty of $4\cdot 10^{-6}$ (to about $3.0\cdot 10^{-6}$). Consequently, neither Jupiter's nor Europa's internal dissipation can be clearly distinguished in the JUICE tracking data alone. However, as was mentioned in Section \ref{sec:astrometryContribution}, the tracking data from any additional missions to the Jovian system will be strongly synergistic with the analysis of the JUICE tracking data.

\subsection{Jupiter ephemeris uncertainty}
\label{sec:jupiterResults}
As a byproduct of our simulations for the ephemeris uncertainty of the Galilean moons, we obtain formal uncertainties for Jupiter's position, as estimated from simulated JUICE tracking data. However, the dynamics of Jupiter itself is influenced by many additional parameters that were not considered there. Most notably, external perturbations due to asteroids significantly influence planetary dynamics, whereas they reduce to weaker tidal terms in the problem of planetary satellite dynamics (in a local frame). We do not include asteroid mass uncertainty or estimation in our simulations. As a result, the formal errors that we obtain for the initial state of Jupiter will be overly optimistic to a greater degree than those of the Galilean satellites.

%However, as discussed in Section \ref{sec:covarianceAnalysis}, these values will be more optimistic than the formal errors of the satellite ephemerides  %as a result of acceleration-model inaccuracies, which are much more influential for Jupiter's barycentric motion than for the local dynamics of the moons 
%(Section \ref{sec:covarianceAnalysis}). Consequently, the \emph{true} error may be significantly higher for Jupiter than what we obtain from our simulations. 

Nevertheless, our results provide preliminary insight into the information content of the Jovian dynamics that is encoded in the tracking data (Tables \ref{tab:absoluteNominalVlbiContribution} and \ref{tab:nominalVlbiContribution}). Also, we note that the use of Gaia astrometric data will allow the masses and dynamics of a number of asteroids to be improved \citep{HestrofferEtAl2010}, in part mitigating dynamical modelling error for Jupiter.

The in-plane component of Jupiter's initial position is determined at 50-200 m formal uncertainty for the solution without VLBI, and 20-100 m for the solution that includes the VLBI data. However, at these levels of uncertainty, it is unclear whether the measurement error will be the dominant true error source for the ephemerides. %, or whether the dynamical model uncertainty limits the data from being fully exploited.   

Without the VLBI data, the out-of-plane component is quite sensitive to the spacecraft's position uncertainty, which ranges from 1,200 m (nominal) to 6,000 m (maximum off-nominal). As was the case for Ganymede's and Callisto's epheme\-rides, the inclusion of the VLBI data greatly reduces the uncertainty of this component, which is decreased to approximately 20 m, 50 m and 80 m for 0.1, 0.5 and 1.0 nrad measurement uncertainties, respectively (for JUICE position uncertainty case 4). The improvement in out-of-plane component is even stronger for Jupiter than for the moons. Less than a single orbit of Jupiter is observed, as opposed to about 175 orbits for Ganymede during JUICE's 3.5 year mission. However, it is likely that the uncertainty when including the VLBI data will be partly damped by dynamical model errors.

Varying the settings of our simulations shows that it is especially the Ganymede VLBI tracking data that contributes to the determination of the Jovian ephemeris, both during the flybys and the orbit phase (Table \ref{tab:planningInfluenceResults}). Both the optical astrometry and Callisto/Europa flyby tracking data contribute relatively little to the Jovian ephemeris.% (Table \ref{tab:opticalAstrometry}). 

Tracking data from the Juno spacecraft \citep{JonesEtAl2017}, which is presently orbiting Jupiter, will greatly improve the current uncertainty in the ephemeris of the Jovian system. Also, the Jupiter VLBI data of the Ulysses spacecraft \citep{FolknerEtAl1996} will aid in constraining Jupiter's long-term dynamics. In addition, the Jovian gravity fields that the Doppler data from the Juno mission will produce \citep{FolknerEtAl2017} ensure that Jovian gravity field uncertainty  will not impact the ephemeris generation of the satellites from JUICE data \citep{DirkxEtAl2016}. However, the Juno mission will likely end about 10 years before JUICE's arrival in the system. Combining the data sets of the two missions will provide crucially accurate input to the generation of solar system ephemerides. %Such a combination should be analyzed in any future analyses of JUICE Jupiter ephemeris quality.

\section{Conclusions}
\label{sec:conclusions}

We have set up a covariance analysis to assess the accuracy to which the ephemerides of the Galilean moons can be determined from  3GM, PRIDE, JANUS and NavCam data from the JUICE mission, with a focus on the relative contribution of the VLBI data produced by PRIDE. Our results indicate that the main contributions of the VLBI data to the generation of the Jovian system ephemeri\-des fall into two categories
\begin{itemize}
\item The strong improvement of the out-of-plane component of the ephemerides, particularly for Ganymede and Callisto.
\item The reduction of the influence of the orbit determination quality of the JUICE spacecraft on the uncertainty of the ephemerides.
\end{itemize}

At many points in the results, we observe distinct behaviour for the results obtained using 0.1 nrad VLBI data on the one hand, and 0.5 and 1.0 nrad data on the other hand. As such, it is strongly recommended that the Ka-band signal be exploited for PRIDE lateral position data, as well as the planned use for 3GM. However, this will require a much denser catalogue of reference sources in this frequency range, as well as a broader availability of Ka-band receiver hardware at the stations. An effort to densify a catalogue of reference sources in the ecliptic plane is currently underway \citep{ShuEtAL2017}. The task is also synergistic to the ongoing effort in creating the next-generation VLBI-based International Celestial Reference Frame, ICRF3, with the aim of providing the reference sources position uncertainty floor at 0.2 nrad \citep{MalkinEtAl2015}.%, which will present definite challenges for the VLBI community. 

Our numerical simulations show that creating epheme\-rides from JUICE data alone will require the inclusion of spacecraft optical astrometry, if the data is to be weighted at its inherent quality. However, even with the inclusion of optical astrometry, the solution of especially Io becomes unstable in certain cases. The combination with other missions in the Jovian system (IVO; Europa Clipper), or optimization of the optical astrometry schedule, will at least partly mitigate this issue.

The estimation uncertainty of Io's $k_{2}/Q$ is largely insensitive to both the planning and uncertainty of the VLBI observations. This is due to the fact that uncertainties in Io's and Europa's dynamics limit the degree to which this parameter can be estimated. Nevertheless, our results indicate a formal error that is similar to that obtained by \cite{LaineyEtAl2009}. %The dissipation inside Jupiter cannot be constrained by JUICE data alone to a level that is similar to the existing estimate from astrometry. 
Also, the estimation is not able to substantially improve upon the \emph{a priori} estimate of the dissipation in either Jupiter or Europa. {However, a combination of JUICE tracking data with Earth-based astrometric and photometric data, as well as data from other missions (Juno, Europa Clipper, \emph{etc.}) will allow an improved determination of dissipation parameters. These improvements in the ephemerides will shed new light on the origin and evolution of the Jovian system, and allow an improved characterization of the stability of the conditions inside the satellites, with key implications to the analysis of icy satellites as habitats. Moreover, improvements made in satellite ephemerides early in the mission will reduce the uncertainties in later flybys, reducing the $\Delta V$ required for trajectory corrections.}

Finally, we have found that the VLBI data could be especially important for the determination of the ephemeris of Jupiter itself. However, the degree to which measurement uncertainties and dynamical model errors will enter the error budget of the ephemeris generation must be investigated in more detail before any robust statements can be made.

In this article, we have taken a broad view of the influence of PRIDE-JUICE on the ephemerides. Our primary model simplification has been the decoupling of the orbit determination and ephemeris generation. Future analyses, in which the dynamics of the moons, Jupiter and the spacecraft itself are concurrently considered will be crucial in determining the contribution that PRIDE's Doppler data will provide, as a supplement to the 3GM data. Additionally, any possible contribution of the VLBI data to the orbit determination, especially in between flybys, where orbit reconstruction is typically less accurate, remains to be investigated. This last point is especially critical in the application of the constrained multi-arc solution strategy.

\section*{Acknowledgements}
The authors are indebted to Luciano Iess for many discussions on the performance of 3GM radio tracking data, and for promptly providing a nominal range data measurement schedule. Pasquale Palumbo is thanked for discussions on JANUS observations. We thank Hermes Jara-Orue for providing a modelled upper bound for Europa's $k_{2}/Q$. {The ESA SPICE team is thanked for providing kernels of the JUICE spacecraft, which can be found at:}\\

 \small \texttt{www.cosmos.esa.int/web/spice/spice-for-juice}

\appendix
\renewcommand{\thesubsection}{\Alph{section}.\arabic{subsection}}

\section{Covariance matrix generation}
\label{app:lsqOd}
In this appendix, we provide a brief overview of the manner in which the covariance matrix is generated in our simulations\citep[\textit{e.g.}][]{MontenbruckGill2000, MilaniGronchi2010}. Let $\mathbf{h}$ denote the set of modelled observations used as input to the analysis and $\mathbf{q}$ the set of parameters that is to be estimated. The design matrix $\mathbf{H}$ is then formed by computing:
\begin{align}
\mathbf{H}=\frac{\partial \mathbf{h}}{\partial \mathbf{q}}
\end{align}
The covariance matrix $\mathbf{P}$ is then computed from the following
\begin{align}
\mathbf{P}&=\left(\mathbf{P}_{0}^{-1}+\left(\mathbf{H}^{T}\mathbf{W}\mathbf{H} \right)\right)^{-1}
\end{align}
where $\mathbf{P}_{0}$ is the \emph{a priori} covariance matrix and $\mathbf{W}$ is the weight matrix of the observations, which we set as a diagonal matrix with $W_{ii}=\sigma_{h,i}^{-2}$, implicitly assuming the measurement uncertainties to be uncorrelated. Here, $\sigma_{h,i}$ denotes the uncertainty of observation $i$ (\emph{e.g.} the observation for which row $i$ of $\mathbf{H}$ denotes its influence on the estimated parameters). 

In the above, $\mathbf{P}$ denotes the covariance matrix, from which the formal errors $\sigma_{q,j}$ of the estimated parameters are computed from the diagonal of the covariance matrix as follows:
\begin{align}
\sigma_{q,j}=\sqrt{P_{jj}}
\end{align}
which are the values of the uncertainties that we discuss in the body of this manuscript.

\section{Comparison with preliminary JUICE orbit determination}
\label{app:giacomo}

In this appendix, we briefly discuss the preliminary results of JUICE orbit determination simulations using the \emph{Orbit14} software \citep{CicaloEtAl2016}, keeping the moons' states fixed. These simulations are meant to validate the assumptions we have made in this article, as they provide (formal) values of the uncertainties we assumed (as listed in  Table \ref{tab:estimatedParameterTable}) .

The values presented as 'Case 1' in Table \ref{tab:estimatedParameterTable} correspond relatively well to the simulated position uncertainties of the spacecraft during the flyby/orbit phase, as obtained from \emph{Orbit14}. We (preliminarily) find an along- and cross-track formal uncertainty of 3-4 m, and a radial formal positon uncertainty of 0.5 m during GEO/GCO5000. Similarly, values of 2.5 m for the along-track component, and about 0.25 m for both the radial and cross-track component, are obtained during GCO500. These values compare reasonable well to the Case 1 we have defined in Table \ref{tab:estimatedParameterTable} for these two cases (with the clear exception of the cross-track component during GCO500). The \emph{true} error will be higher than a simulated formal error, especially for spacecraft orbit determination (Section \ref{sec:covarianceAnalysis}). For the tracking data analysis from JUICE, though, this true-to-formal error ratio may not be as high as for many past and current missions due to the use of the dual-band tracking system. This system will cancel a substantial part of the time-correlated noise. Based on these results, we conclude that our assumed spacecraft orbit uncertainties for GCO500/GEO5000/GCO5000 lie within a realistic range, although the 'Case 5' may very well be too pessimistic. However, we note that the marginal position uncertainty of the moons will be substantially larger than their conditional uncertainty (Section \ref{sec:juiceDynamics}).

Formal orbit uncertainties during the Europa flybys for the JUICE spacecraft are about a factor 4 better than our 'Case 1' uncertainties. The exceptional quality of the orbit reconstruction at Europa should be verified by more detailed analyses of the influence of various (non-Gaussian) error sources, but could be instrumental in reducing the ill-posedness of the ephemeris solution. The results for the JUICE formal estimation error at Callisto are reasonably well in line with Case 1 for about half of the flybys. For the rest of the flybys, the geometry of the spacecraft orbital plane w.r.t. the Earth is very close to edge-on {(see Fig. \ref{fig:flybyGeometry})}, reducing the signature of the cross-track dynamics on the Doppler tracking data. Consequently, uncertainties of $>100$ m in cross-track direction are obtained for these flybys. As the influence of the Callisto tracking data is largely limited to the estimation of Callisto itself, this effect will likely somewhat worsen its ephemeris uncertainty, but will not have substantial consequences for the other moons. A similar effect is observed for Ganymede, where the tracking geometry is poor for two flybys. As this geometric effect does not impact the first several flybys or the orbit phase, the impact on the total ephemeris reconstruction will likely remain limited, though. For these particular flybys, the use of VLBI data, which is sensitive to the cross-track dynamics at edge-on geometry, may be valuable.

Finally, we note that the ratio of components of $\boldsymbol{\sigma}_{{\mathbf{r}}_{sc}^{i}}$ and $\boldsymbol{\sigma}_{\dot{\mathbf{r}}_{sc}^{i}}$, which we use to approximate spacecraft velocity uncertainties, consistently lie within $10-20 \%$ of the values obtained with \emph{Orbit14}.

The analysis of the influence of the exact flyby geometry should be treated with caution, as the precise geometry of the JUICE trajectory may very well be substantially different from the current working orbit. Although the total number of flybys, as well as their approximate times in the overall mission, are unlikely to be strongly modified at this point,  modifications may result in a given flyby having a face-on, instead of edge-on, geometry (or \emph{vice versa}).

\bibliographystyle{apalike} 
\bibliography{Bibliography/Bibliography}

\end{document}